\documentclass[twocolumn,showpacs,amsmath,amssymb]{revtex4}

\usepackage[dvips]{epsfig}
\usepackage[dvips]{graphics}

\newcommand{\oper}{\mathcal{O}}

\newcommand{\black}[1]{\mathbf{#1}}
\newcommand{\base}[1]{\widehat{\black{e}}_{#1}}

\newcommand{\firstG}{\mathcal{G}_1}
\newcommand{\secondG}{\mathcal{G}_2}
\newcommand{\thirdG}{\mathcal{G}_3}
\newcommand{\fourthG}{\mathcal{G}_4}
\newcommand{\fifthG}{\mathcal{G}_5}
\newcommand{\sixthG}{\mathcal{G}_6}

\begin{document}

\title{Topological signatures in CMB temperature anisotropy maps}

\author{W.S. Hip\'olito--Ricaldi}
 \email{hipolito@ift.unesp.br}
\author{G.I. Gomero}%
 \email{german@ift.unesp.br}
\affiliation{%
Instituto de F\'{\i}sica Te\'orica, \\
Universidade Estadual Paulista,  \\
Rua Pamplona 145 \\
S\~ao Paulo, SP 01405--900, Brazil
}

\date{\today}

\begin{abstract}
We propose an alternative formalism to simulate CMB temperature maps in
$\Lambda$CDM universes with nontrivial spatial topologies. This formalism
avoids the need to explicitly compute the eigenmodes of the Laplacian operator
in the spatial sections. Instead, the covariance matrix of the coefficients of
the spherical harmonic decomposition of the temperature anisotropies is
expressed in terms of the elements of the covering group of the space. We
obtain a decomposition of the correlation matrix that isolates the topological
contribution to the CMB temperature anisotropies out of the simply connected
contribution. A further decomposition of the topological signature of the
correlation matrix for an arbitrary topology allows us to compute it in terms
of correlation matrices corresponding to simpler topologies, for which closed
quadrature formulae might be derived. We also use this decomposition to show
that CMB temperature maps of (not too large) multiply connected universes must
show ``patterns of alignment'', and propose a method to look for these
patterns, thus opening the door to the development of new methods for
detecting the topology of our Universe even when the injectivity radius of
space is slightly larger than the radius of the last scattering surface. We
illustrate all these features with the simplest examples, those of flat
homogeneous manifolds, i.e., tori, with special attention given to the
cylinder, i.e., $T^1$ topology.
\end{abstract}


\maketitle

%%%%%%%%%%%%%%%%%%%%%%%%%%%%%%%%%%%%%%%%%%%%%%%%%%%%%%%%%%%%%%%%%%%%%%%%%%%%%%%
\section{Introduction}
\label{S:Intro}
%%%%%%%%%%%%%%%%%%%%%%%%%%%%%%%%%%%%%%%%%%%%%%%%%%%%%%%%%%%%%%%%%%%%%%%%%%%%%%%

It is becoming widely recognized that universes with a nontrivial spatial
topology may be more natural models for our Universe than the traditional
simply connected ones. This naturalness can be invoked from the mathematical
point of view by arguing that there is an infinity of locally homogeneous and
isotropic multiply connected 3--spaces, while there are only three simply
connected ones; or with physical arguments coming from incursions into the
nobody's territories of Quantum Gravity and Quantum Cosmology. On the other
hand, from a more pragmatic point of view, we can argue that cosmological
models with a nontrivial spatial topology offer a very rich field of research,
and are particularly well suited to explain certain reported ``anomalous''
features in CMB temperature maps, such as the alignments of their low
$\ell$--modes \cite{Aligne1, Aligne2}.

Conversely, the full sky CMB temperature maps produced by the space missions
COBE and WMAP provide us with an amazingly rich and high quality amount of
data with which we can look for the topology of space. This is very compelling
for those who wish to unmask our Universe and see its shape, since Cosmic
Topology is at present an almost exclusively observational and
phenomenological issue, due to the lack of an accepted fundamental physical
theory which can predict the global topology of space.

Theory demands topology of space to leave several different kinds of marks in
CMB temperature maps. Two of them have been largely studied and exploited to
try to unveil the shape of our Universe, the distorsion of the angular power
spectrum with respect to that of a simply connected universe
\cite{PowSpec}--\cite{Weeks04}, and the existence of  ``circles in the sky''
\cite{CinSky, CSSK04}. Two other closely related signatures, a non--null
bipolar power spectrum \cite{StatAnis, SH05}, and alignments of the low
$\ell$--modes \cite{Aligne1, Aligne2, Weeks04}, have been only marginally
used. Our main motivation for deciding to adventure into cosmic topology with
the CMB was the desire to get a deeper understanding of the nature and
properties of these alignments as a topological signature.

One indispensable tool for a project like this is a software facility to
produce simulated CMB temperature maps in multiply connected universes, so
that we could systematically study the effects of different sizes and topologies on
these alignments. These simulation procedures exist and have been used in
several studies in cosmic topology \cite{Aurich}, \cite{LevinB}--\cite{BPS},
so we could expect that this issue of the project would not present any
problem.

However, almost all known methods for computing CMB temperature anisotropies
in multiply connected universes need to solve the Helmholtz equation in the
manifold modelling our 3--dimensional space, the only exception to our
knowledge being the work of Bond et al. \cite{BPS}. To solve the Helmholtz
equation is a relatively easy problem in Euclidean manifolds \cite{LevinA,
RWULL04}, but a very difficult task in spherical \cite{SphEigen, Lachieze} and
hyperbolic 3--spaces \cite{HypEigen}. Indeed, the spherical case has been
completely solved analitically only recently by Lachi\`eze--Rey
\cite{Lachieze}, while for compact hyperbolic manifolds the only possible
approach is numerical \cite{HypEigen}.

Among other results, in this paper we present a new approach to the
computation of the correlation matrix $\langle a_{\ell m} a_{\ell' m'}^*
\rangle$ of the coefficients of the spherical harmonic decomposition of CMB
temperature anisotropies in a universe with nontrivial spatial topology. The
main feature of this approach is that it avoids the explicit computation of
the solutions of the Helmholtz equation in the spatial sections of spacetime.
Instead, we express the correlation matrix in terms of the covering group
alone. Incidentally, the idea of generating a CMB map exploiting the
symmetries of the quotient space was already suggested time ago by Janna Levin
and collaborators \cite{LevinB, LSS98}. In particular we wish to quote a
citation of a nontrivial claim in \cite{LSS98} (p.2695) which we have, in our
opinion, succeeded in achieving: ``By understanding the symmetries of the
fundamental polyhedron and the identification rules, a CBR pattern can be
deduced without the need to explicitly obtain the spectrum mode by mode.''

Our main formal result is a generic decomposition of the form
\begin{equation}
\label{GenDecomp}
X^\Gamma = X^{s.c.} + X^{t.s.} \; ,
\end{equation}
where $X$ may be any covariance function which can be related to the
two--point correlation function of the Newtonian potential (see
Sec. \ref{S:CycDec}), $\Gamma$ is the covering group of (the multiply
connected) space, $s.c.$ stands for ``simply connected'', and $t.s.$ means
``topological signature''.

Thus, in Eq. (\ref{GenDecomp}), $X^\Gamma$ is the covariance function computed
in the manifold $M=\widetilde{M}/\Gamma$, and $X^{s.c.}$ the same covariance
function but computed in the universal covering space $\widetilde{M}$. It
means that all the topological information is encoded in the ``perturbative''
term $X^{t.s.}$, and that is why we refer to it, generically, as the
topological signature of $X^\Gamma$. This decomposition is always possible
whenever one can express $X^\Gamma$ in terms of $\Gamma$, as for example in
the PSH method for detecting multiple copies of standard candles
\cite{CosCris}. In the present case we succeeded in writing the correlation
matrix of the $a_{\ell m}$'s in this way by formally manipulating the
two--point correlation function of the Newtonian potential derived by Bond,
Pogosyan and Souradeep \cite{BPS}.

Our approach to compute the correlation matrix $\langle a_{\ell m} a_{\ell'
m'}^* \rangle$ has some methodological advantages in the simulation of CMB
temperature maps. Indeed, by means of a suitable decomposition of the covering
group $\Gamma$ in cyclic subgroups, we are able to write down a formula for the
correlation matrix of a complicated topology in terms of the correlation
matrices of the cyclic topologies (topologies with a cyclic covering group)
that cover it maximally. Since correlation matrices for cyclic manifolds are
relatively easy to compute (we obtain a closed quadrature formula for the
cylinder), we expect to obtain in the near future more efficient ways to
simulate maps for complicated manifolds.

The decomposition of $\Gamma$ in cyclic subgroups describes in a transparent
way the symmetries of the manifold, and this fact gives rise to another
advantage of our approach, in this case from the observational side. Universes
with a cyclic topology present an alignment along the ``direction of the
generator isometry''. It follows that a CMB map in a universe with a
nontrivial topology might present ``patterns of alignment'' (one for each
$\ell$--mode) characteristic of its shape and size. Indeed, and this shall
become ``obvious'', symmetries of the quotient manifolds translate into
symmetries of their patterns of alignment. We propose a method to search for
these patterns by constructing maps of the dispersion of the squares
$|a_{\ell m}|^2$ around the power spectrum. This opens the door to the
development of methods to look for topology by searching these patterns,
instead of limiting ourselves to considerations concerning only the special
directions defined by the alignments.

For the sake of brevity, other advantages of our approach to compute the matrix
$\langle a_{\ell m} a_{\ell' m'}^* \rangle$ are discussed in
Sec. \ref{S:Discuss} only. We prefer now to make a few remarks on some
limitations of our work. We have considered here a few simplifications to
develop the formalism, and worked out the details for the very simplest
nontrivial topologies. In fact, we (i) have considered the Sachs--Wolfe effect
as the only source of temperature anisotropies in CMB maps, (ii) have written
the correlation matrix $\langle a_{\ell m} a_{\ell' m'}^* \rangle$ formally
only for Euclidean 3-spaces, (iii) have worked out the details for homogeneous
flat manifolds, (iv) have made a detailed analysis and some simulations only
for cylinders ($T^1$ topology), and (v) these simulations were done using the
Einstein--de Sitter model. We wish to close this introduction by justifying
each one of these simplifications.

The shape of space is a global property, thus we expect the topological
signatures in CMB to show themselves on very large scales only. Although a
proof is missing, we believe that the main features of these signals would
observationally appear if we restrict the searches to the low $\ell$--modes in
the temperature maps, i.e., we do not need high resolution CMB maps in Cosmic
Topology! Since the Sachs--Wolfe effect is the main source of temperature
anisotropies at these scales \cite{RULW04}, we expect that theoretical
explorations considering only this effect will put in evidence the main
features of the topological signals that we would observe in a real map. The
addition of the missing part of the anisotropies will only modify
quantitatively the predictions made with our approximation, and thus, will
only be important when adjusting theoretical models with data.

We consider this paper technically hard, so much care has been taken to write
it in a clear and pedagogical way. The main features of our formalism and of
the topological signatures we predict in CMB can be understood by restricting
the presentation to Euclidean topologies. The inclusion of nonzero spatial
curvature will only introduce additional technical considerations (and nothing
qualitatively new), thus we decided to leave the spherical and hyperbolic
cosmological models for a future paper.

The same applies to the lack in the paper of detailed explicit correlation
matrices for nonhomogeneous flat manifolds. Indeed, the price we pay for
simplicity and transparency in the presentation of the results for each
specific cyclic topology is the need for very hard calculations in the middle
steps, as can be seen in the appendices. Explicit calculations for
nonhomogeneous flat manifolds will only add one page to the main body of the
paper, and one or two more appendices to the already large list of them (see
the end of the introduction). An exhaustive presentation for all the Euclidean
manifolds is left for a future work.

The last but not the least, we performed the simulations with the Einstein--de 
Sitter model for simplicity. Nothing is lost from the theoretical point of
view with this simplification since, as discussed in the paper, the structure
of the topological signatures in CMB is captured in this oversimplified and old
fashioned model of our Universe. However, more realistic $\Lambda$CDM models
will be required to confront theory with observations quantitatively.

We close this introduction by giving a detailed description of the structure of
the paper. In Sec. \ref{S:CMBSimul} we briefly review the two most common
methods to simulate CMB temperature maps in multiply connected universes, as
well as present the method we have developed, and implemented for the
Euclidean case. In Sec. \ref{S:CycDec} we define the topological signature in
a correlation function, perform the decomposition of the covering group of a
quotient space in its cyclic subgroups, and write the topological signature in
terms of this decomposition. We also show here that the symmetries of a
quotient space appear transparently in the decomposition of its covering group
in cyclic subgroups.

In Sec. \ref{S:tori} we apply our formalism to the homogeneous Euclidean
manifolds, which are the simplest. We first compute the correlation matrix and
the angular power spectrum for the cylinder, and apply the general results of
the previous section in order to write down the correlation matrix of the
spherical harmonic coefficients and the angular power spectrum for a generic
torus. Finally, as an illustration, we write those expressions explicitly for
the chimney ($T^2$ topology).

In Sec. \ref{S:TopSign} we first show, by means of simulations, that in a
universe with the topology of a cylinder the low $\ell$--modes are aligned in
a similar fashion as they are in the WMAP data. We then use the results in the
previous sections to argue that CMB temperature maps in a universe with a
nontrivial topology must present characteristic patterns of alignment, and
propose the method of mapping on the sphere the dispersion of the squares
$|a_{\ell m}|^2$ to look for them. Finally, in Sec. \ref{S:Discuss} we discuss
in detail the results of this paper and suggest further lines of research.

In brief, the main goal of this paper is to show that our approach to the
computation of the correlation matrix $\langle a_{\ell m} a_{\ell' m'}^*
\rangle$ (Sec. \ref{S:CMBSimul}), together with the decomposition of the
covering group of a manifold in cyclic subgroups (Sec. \ref{S:CycDec}), led
to the discovery of a new topological signature in CMB temperature maps, i.e.,
the ``patterns of alignment'' (Sec. \ref{S:TopSign}). To illustrate this we
have used the simplest example, i.e., that of flat homogeneous manifolds (Sec.
\ref{S:tori}).

The paper has four appendices. In Appendix \ref{Ap:SphHarm} we collect
standard definitions and results related to spherical harmonics in order to
set the conventions used in this paper. The technical calculations needed for
the computation of the correlation matrix $\langle a_{\ell m} a_{\ell' m'}^*
\rangle$ for the cylinder are presented in Appendices \ref{Ap:Clausen} and
\ref{Ap:EvalFllm}. In the former we develop from scratch the theory of Clausen
$\varphi$--functions, for which we have not found any suitable reference in
the literature. In the latter we compute a function that is the key part for
computing efficiently the topological signature of the correlation matrix for
the cylinder. Finally, in Appendix \ref{S:OldRes} we briefly reproduce known
results for the correlation matrix of five out of the six compact orientable
Euclidean 3--spaces. These formulae have been obtained previously in the
literature by considering explicitly the solutions of the Helmholtz equation
in these manifolds, and are written in terms of the $k$--modes \cite{LevinA,
RWULL04}. Our derivation avoids the need for considering these solutions.

%%%%%%%%%%%%%%%%%%%%%%%%%%%%%%%%%%%%%%%%%%%%%%%%%%%%%%%%%%%%%%%%%%%%%%%%%%%%%%%
\section{Simulating CMB temperature maps}
\label{S:CMBSimul}
%%%%%%%%%%%%%%%%%%%%%%%%%%%%%%%%%%%%%%%%%%%%%%%%%%%%%%%%%%%%%%%%%%%%%%%%%%%%%%%

In this section we briefly describe two methods currently available for
simulating CMB temperature maps in universes with non--trivial spatial
topology, and proceed to develop our own formulation. We consider $\Lambda$CDM
universes, where the background metric of spacetime is of the
Robertson--Walker type, and include scalar and adiabatic perturbations as the
seeds for the temperature anisotropies of the CMB. In the Newtonian gauge we
have
\[
ds^2 = a^2(\eta) \left[(1 + 2\Phi)d\eta^2 - (1 - 2\Phi)\gamma_{ij} dx^i dx^j
\right]
\]
for the metric, where $\eta$ is the conformal time, $a(\eta)$ is the scale
factor,  $\Phi$ is the Newtonian potential, and
\[
\gamma_{ij} = \left( 1 + \frac{K}{4} (x^2 + y^2 + z^2) \right)^{-2}
\delta_{ij}
\]
is the metric of the spatial section of the background with sectional
curvature $K=0$, $\pm 1$.

The matter content consists of radiation ($\Omega_r$), baryonic and cold dark
matter ($\Omega_m = \Omega_b + \Omega_{cdm}$), and dark energy in the form
of a cosmological constant ($\Omega_{\Lambda}$). Since we are interested on
fluctuations on large angular scales, we make the assumption of instantaneous
recombination and do not consider finite thickness effects. The main
contribution to the temperature anisotropy observed at the direction
$\black{n}$ comes from the complete (ordinary plus integrated) Sachs--Wolfe
effect
\begin{eqnarray}
\label{SW}
\frac{\delta T}{T} (\black{n}) & = & \frac{1}{3} \, \Phi(\eta_{LSS},R_{LSS}
\black{n}) + \\
& & \hspace{1.5cm} 2 \int_{\eta_{LSS}}^{\eta_0} d\eta \left. \frac{\partial
\Phi}{\partial \eta} \right|_{(\eta,R(\eta)\black{n})} \; , \nonumber
\end{eqnarray}
where the index $LSS$ stands for `last scattering surface', the index $0$ for
present time, and $R(\eta)$ is the comoving distance at instant $\eta$ between
a photon, scattered at $\eta_{LSS}$, and the observer.

The Newtonian potential is written as
\begin{equation}
\label{SepVar}
\Phi(\eta,\black{x}) = \int dq F(\eta,q) \xi(q,\black{x}) \; .
\end{equation}
The temporal part satisfies the equation
%\begin{widetext}
\begin{eqnarray}
\label{Temp}
F''(\eta) + 3 \mathcal{H} \left( 1+c_s^2 \right) F'(\eta) + \left[ 2
\mathcal{H}' + \right. \hspace{1cm} & & \\
\left. \left(1+3c_s^2 \right) (\mathcal{H}^2 - K) + c_s^2 q^2 \right] F(\eta)
& = & 0 \; , \nonumber
\end{eqnarray}
%\end{widetext}
where $c_s$ is the speed of sound in the fluid and $\mathcal{H} = a'/a$ is the
Hubble parameter in conformal time. On the other hand, the spatial part
consists of solutions of the Helmholtz equation
\begin{equation}
\label{Helmholtz}
(\Delta + q^2) \, \xi(q,\black{x}) = 0 \; ,
\end{equation}
where the index $q$ has been put as a variable in $\xi$ for simplicity of
notation.

The integral in Eq. (\ref{SepVar}) has to be understood in a measure theoretic
sense. Indeed, for multiply connected spaces the measure $dq$ is not the usual
one but a combination of a discrete and an absolutely continuous measures,
reducing the integral in (\ref{SepVar}) to a sum and an integral in the usual
sense. In particular, if the space is compact, the measure reduces to a
discrete one. This comes from the well--known fact that not every eigenmode of
the Laplacian operator in the universal covering space $\widetilde{M}$ is also
an eigenmode in a quotient space $M = \widetilde{M}/\Gamma$. In fact, only
eigenmodes in $\widetilde{M}$ satisfying the invariance conditions
\begin{equation}
\label{automorph}
\xi(q,g \black{x}) = \xi(q,\black{x})
\end{equation}
for any $g \in \Gamma$ project to eigenmodes in $M$.

The most straightforward way of simulating CMB temperature maps is by solving
(\ref{Temp}) and (\ref{Helmholtz}), performing the sum in (\ref{SepVar}), and
then evaluating the SW effect (\ref{SW}). However, one has to consider that a
temperature anisotropy map is a realization of a random field on the
2--sphere, and this randomness is inherited from that of the Newtonian
potential (\ref{SepVar}). There are currently two ways to implement this
random character in the simulations, and one goal of this paper is to propose
a third one.

The first and most direct method is to consider the randomness in the temporal
part of the decomposition (\ref{SepVar}) of the Newtonian potential. The
two--point correlation function of the Newtonian potential at fixed time
$\eta$ can then be written as
\begin{eqnarray}
\label{twopointAur}
\langle \Phi(\eta,\black{x}) \, \Phi(\eta,\black{x}') \rangle & = & \int dq \,
dq' f(\eta,q,q') \times \\
& & \hspace{1.2cm} \xi(q,\black{x}) \, \xi^*(q',\black{x}') \; , \nonumber
\end{eqnarray}
where $f(\eta,q,q') = \langle F(\eta,q) \, F(\eta,q') \rangle$ is the
two--point correlation function for the amplitudes of the scalar perturbation
modes and $\xi(q,\black{x})$ are normalized solutions of the Helmholtz
equation.

Assuming that the Newtonian potential is a homogeneous and isotropic random
field, the two point correlation function (\ref{twopointAur}) reduces to a
function of time $\eta$ and the distance $d(\black{x}, \black{x}')$, and thus
we get $f(\eta,q,q') = P_\Phi(\eta,q) \, \delta(q-q')$, where $P_\Phi(\eta,q)$
is the gravitational power spectrum. If, in addition, the Newtonian potential
is assumed to be gaussian, its random character is completely encoded in the
variance of the temporal part
\begin{equation}
\label{VarianceTemp}
\langle F^2(\eta,q) \rangle = P_\Phi(\eta,q) \; .
\end{equation}

Specifying this function, one then takes as an initial condition,
$F(\eta_{init},q)$, a realization of a normal distribution with zero mean and
variance given by Eq. (\ref{VarianceTemp}), and some suitable condition for
the initial first derivative. With these initial conditions one solves for
(\ref{Temp}), so one can now compute the potential (\ref{SepVar}). Topology is
considered by restricting in (\ref{SepVar}) to normalized solutions of the
Helmholtz equation satisfying the invariance conditions (\ref{automorph}).
This method has been extensively used in \cite{Aurich, LevinB, LSS98},
although in the former the authors do not consider the randomness of the
function $F(\eta,q)$. Instead the random character of the CMB maps is
attributed exclusively to the random character of the eigenmodes of the
Laplacian in compact hyperbolic spaces.

The second method to produce simulated maps of CMB temperature anisotropies in
universes with nontrivial spatial topology was first described in
\cite{RULW04}, and used in \cite{URLW04, RWULL04}. It is based in considering
the randomness of the Newtonian potential in the eigenmodes of the Laplacian
operator. We begin by decomposing the general solution of the Helmholtz
equation, in the universal covering space, as a sum of fundamental solutions
\begin{equation}
\label{HelmGen}
\xi(q,\black{x}) = \sum_{\ell,m} \widehat{\xi}_{\ell m}(q)
\mathcal{Y}_{\ell m}(q,\black{x}) \; ,
\end{equation}
where
\begin{equation}
\label{HelmRadAng}
\mathcal{Y}_{\ell m}(q,\black{x}) = \rho_\ell(q,x) \, Y_{\ell m}(\black{n})
\end{equation}
is the normalized solution of the Helmholtz equation after separating it in
radial and angular variables. Here we have put $x = |\black{x}|$, $\black{n}$
is the unit vector in the direction of $\black{x}$, and the $Y_{\ell
m}(\black{n})$ are the spherical harmonic functions (see appendix
\ref{Ap:SphHarm}). Since the solutions (\ref{HelmRadAng}) are normalized, the
randomness of the eigenmodes' amplitudes relies on the coefficients
$\widehat{\xi}_{\ell m}(q)$.

Introducing (\ref{SepVar}) in (\ref{SW}), and using (\ref{HelmGen}) and
(\ref{HelmRadAng}), we arrive at the decomposition of the temperature
anisotropy map in spherical harmonics
\begin{equation}
\label{TempSphHarm}
\frac{\delta T}{T} (\black{n}) = \sum_{\ell,m} a_{\ell m} Y_{\ell m}(\black{n}) \; ,
\end{equation}
with multipole coefficients
\begin{equation}
\label{HarmCoeff}
a_{\ell m} = \int dq \, \widehat{\xi}_{\ell m}(q) \, G_\ell(q) \; ,
\end{equation}
and the effects of physical cosmology given by
\begin{eqnarray}
\label{Gl(q)}
G_\ell(q) & = & \frac{1}{3} F(\eta_{LSS},q) \, \rho_\ell(q,R_{LSS}) + \\
& & \hspace{1cm} 2 \int_{\eta_{LSS}}^{\eta_0} \hspace{-0.2cm} d\eta \left.
\frac{\partial F}{\partial \eta} \right|_{(\eta,q)} \! \rho_\ell(q,R(\eta))
\; . \nonumber
\end{eqnarray}

At this point it is convenient to recall how does topology enter in the story.
Note that, due to the invariance conditions (\ref{automorph}), not every
solution of the form (\ref{HelmGen}) is a solution in a quotient space.
However, one would expect that any solution in a quotient space could be
written in this form, the only problem being to find the correct coefficients
$\widehat{\xi}_{\ell m}(q)$. These coefficients are not independent one from
the other, since the invariance conditions (\ref{automorph}) establish certain
relations among them. In what follows we will assume that these relations can
always be found, so that we will always represent an eigenmode in a quotient
space by Eq. (\ref{HelmGen}), with suitable coefficients.

A crucial point in \cite{RULW04} is the decomposition of these coefficients as
\begin{equation}
\label{Xi_Decomp}
\widehat{\xi}_{\ell m}(q) = \sqrt{P_\Phi(q)} \, \base{\ell m}(q) \; ,
\end{equation}
where $P_\Phi(q)$ is the gravitational initial power spectrum, and the
$\base{\ell m}(q)$ form a multivariate gaussian random variable, with a
non--diagonal covariance matrix due to the relations among the coefficients
$\widehat{\xi}_{\ell m}(q)$ coming from the invariance conditions.

The simulation procedure can now be described. First we solve Eq. (\ref{Temp})
using the initial condition $F(\eta_{init},q) = 1$, and a suitable condition
for the first derivative, and use this in (\ref{Gl(q)}) to compute
$G_\ell(q)$. Then generate a realization of the random variable
$\base{\ell m}(q)$ and use (\ref{Xi_Decomp}) to compute
$\widehat{\xi}_{\ell m}(q)$. The map is now simulated by computing the
coefficients $a_{\ell m}$ using (\ref{HarmCoeff}), and performing the sum in
(\ref{TempSphHarm}).

An alternative method of simulation, also proposed in \cite{RULW04}, is to
construct the covariance matrix of the $a_{\ell m}$'s as
\begin{eqnarray}
\label{CovalmGen}
\langle a_{\ell m} \, a^*_{\ell' m'} \rangle & = & \int dq \, dq' G_\ell(q) \,
G_{\ell'}(q') \times \\
& & \hspace{2cm} \langle \widehat{\xi}_{\ell m}(q) \, \widehat{\xi}^*_{\ell'
m'}(q') \rangle \; . \nonumber
\end{eqnarray}
The substitution of (\ref{Xi_Decomp}) into (\ref{CovalmGen}), and the
evaluation of the resulting integral give rise to expressions for the
covariance matrix in terms of the eigenvalues and eigenmodes of the Laplacian
operator. The multipolar coefficients are then obtained directly as a
realization of a gaussian distribution with zero mean and covariance given by
(\ref{CovalmGen}).

The method we propose in this paper lies along these lines, but we are able to
manipulate the correlation function for the $\widehat{\xi}_{\ell m}(q)$ in a way
that avoids the need for an explicit determination of the eigenmodes of the
Laplacian. Instead, the final expression after the integration of
(\ref{CovalmGen}) is given in terms of the isometries of the corresponding
covering group.

Our starting point is an expression, obtained by Bond et al. in \cite{BPS},
that relates the two--point correlation function of the Newtonian potential in
a simply connected universe, and that in a multiply connected universe, when
both potentials have the same initial power spectrum. For a homogeneous and
isotropic random field the expression is
\begin{equation}
\label{two-point}
\langle \Phi(\eta,\black{x}) \Phi(\eta,\black{x}') \rangle^{\Gamma} =
\sum_{g \in \Gamma} |g| \, \langle \Phi(\eta,\black{x}) \Phi(\eta,g
\black{x}') \rangle^{s.c.} \; ,
\end{equation}
where $|g| =1$ if $g$ is orientation preserving, and $-1$ otherwise.

We now show how to use Eq. (\ref{two-point}) in order to express
(\ref{CovalmGen}) in terms of the covering group, and for simplicity we will
restrict the presentation to flat topologies. In Euclidean space, the most
general solution of Eq. (\ref{Helmholtz}) is written in the form
\begin{equation}
\label{SolHelmEuc}
\xi(q,\black{x}) = \int d^3 k \, \delta(q-k) \, \widehat{\xi}(\black{k}) \,
e^{i \black{k} \cdot \black{x}} \; .
\end{equation}
If we now expand the plane waves in spherical harmonics as
\begin{equation}
\label{PlaneW}
e^{i \black{k} \cdot \black{x}} = 4 \pi \sum_{\ell,m} i^\ell j_\ell(kx)
Y_{\ell m}^*(\black{n}_{\black{k}}) Y_{\ell m}(\black{n}) \; ,
\end{equation}
where $j_\ell(x)$ is the spherical Bessel function of order $\ell$, and
introduce it in (\ref{SolHelmEuc}) we obtain $\xi(q,\black{x})$ expressed as
in Eq. (\ref{HelmGen}) with
\begin{equation}
\label{CoefHelm}
\widehat{\xi}_{\ell m}(q) = 4 \pi i^\ell \int \! d^3k \, \delta(q-k) \,
\widehat{\xi}(\black{k}) Y_{\ell m}^*(\black{n}_{\black{k}}) \; ,
\end{equation}
where $\black{n}_{\black{k}}$ is the unit vector in the direction of
$\black{k}$, and $\rho_\ell(q,x) = j_\ell(qx)$.

Note that we have not decomposed $\widehat{\xi}_{\ell m}(q)$ as in Eq.
(\ref{Xi_Decomp}). Instead, the decomposition (\ref{CoefHelm}) allows us to
implement the randomness of the Newtonian potential in the modes
$\widehat{\xi}(\black{k})$. In fact, introducing (\ref{CoefHelm}) in
(\ref{CovalmGen}), the covariance matrix for the $a_{\ell m}$'s now reads
\begin{eqnarray}
\label{CovalmEucGen}
\langle a_{\ell m} \, a^*_{\ell' m'} \rangle & = & (4 \pi)^2 i^{\ell - \ell'}
\hspace{-0.2cm} \int \! d^3k \, d^3k' G_\ell(k) G_{\ell'}(k') \times
\nonumber \\
& & \hspace{0.2cm} \langle \widehat{\xi}(\black{k}) \,
\widehat{\xi}^*(\black{k'}) \rangle Y_{\ell m}^*(\black{n}_{\black{k}})
Y_{\ell' m'}(\black{n}_{\black{k'}}) \; . \nonumber \\
& &
\end{eqnarray}
It is the correlation function of the modes $\widehat{\xi}(\black{k})$ that
carries all the topological information, as we will see in the following.

Introducing (\ref{SolHelmEuc}) in (\ref{SepVar}) we obtain
\[
\Phi(\eta, \black{x}) = \int d^3k \, F(\eta,k) \, \widehat{\xi}(\black{k}) \,
e^{i \black{k} \cdot \black{x}} \; ,
\]
thus the two--point correlation function of the Newtonian potential now reads
\begin{eqnarray}
\label{TwoPointEuc}
\langle \Phi(\eta,\black{x}) \, \Phi(\eta,\black{x}') \rangle & = & \int d^3k
\, d^3k' F(\eta,k) \, F(\eta,k') \times \nonumber \\
& & \langle \widehat{\xi}(\black{k}) \,
\widehat{\xi}^*(\black{k'}) \rangle \, e^{i (\black{k} \cdot \black{x} -
\black{k'} \cdot \black{x'})} \; .
\end{eqnarray}

At this point we have to recall that an Euclidean isometry can always be
written as $g = (R,\black{r})$, where $R$ is an orthogonal transformation and
$\black{r}$ is an Euclidean vector, and that this isometry acts on a vector
$\black{x}$ as $g \black{x} = R \black{x} + \black{r}$. It is now an easy task
to deduce from Eqs. (\ref{two-point}) and (\ref{TwoPointEuc}) that
\begin{equation}
\label{twopointEucGen}
\langle \widehat{\xi}(\black{k}) \, \widehat{\xi}^*(\black{k'})
\rangle^{\Gamma} = \sum_{g \in \Gamma} \langle \widehat{\xi}(\black{k}) \,
\widehat{\xi}^*(R\black{k'}) \rangle^{s.c} \, e^{-i R\black{k'} \cdot
\black{r}} \; .
\end{equation}

In most inflationary models the initial perturbations of the gravitational
field are homogeneous and isotropic Gaussian random fields, thus the
correlation matrix of the $\black{k}$--modes in a simply connected universe
takes the form
\[
\langle \widehat{\xi}(\black{k}) \widehat{\xi}^*(\black{k'}) \rangle^{s.c} =
\frac{P_{\Phi}(k)}{k^3} \, \delta(\black{k} - \black{k}') \; .
\]
The use of (\ref{twopointEucGen}) now yields
\[
\langle \widehat{\xi}(\black{k}) \widehat{\xi}^*(\black{k'}) \rangle^\Gamma =
\frac{P_{\Phi}(k)}{k^3} \, \sum_{g \in \Gamma} \delta(\black{k} - R\black{k}')
\, e^{-i R\black{k}' \cdot \black{r}} \; ,
\]
for the correlation matrix of the $\black{k}$--modes in the quotient space $M
= \widetilde{M}/\Gamma$, which when substituted in (\ref{CovalmEucGen})
finally gives
\begin{eqnarray}
\label{MCCorr}
\langle a_{\ell m} \, a^*_{\ell' m'} \rangle^{\Gamma} & = & (4 \pi)^2 \,
i^{\ell - \ell'} \int \! \frac{d^3k}{k^3} \, \Psi_{\ell \ell'}(k) \times \\
& & \hspace{2cm} \Upsilon^{\Gamma}_{\ell' m'}(\black{k}) \,
Y_{\ell m}^*(\black{n}_{\black{k}}) \; , \nonumber
\end{eqnarray}
where the physical effects are encoded in
\begin{equation}
\label{PhysSign}
\Psi_{\ell \ell'}(k) = P_{\Phi}(k) \, G_\ell(k) \, G_{\ell'}(k) \; ,
\end{equation}
and the topological information in
\begin{equation}
\label{DefTopSign}
\Upsilon^{\Gamma}_{\ell m}(\black{k}) = \sum_{g \in \Gamma} e^{-i
\black{k} \cdot \black{r}} \, Y_{\ell m}(\black{n}_{R^T\black{k}}) \; .
\end{equation}

The integration in (\ref{MCCorr}) is over the whole $\black{k}$--space. The
topological information is carried in Eq. (\ref{DefTopSign}), which
automatically selects the eigenvalues of the Laplacian operator in $M$. This
can be seen in Appendix \ref{S:OldRes}, where $\Upsilon^{\Gamma}_{\ell
m}(\black{k})$ is expressed in terms of Dirac's delta functions centered in
the eigenvalues of the Laplacian operator in the corresponding quotient
spaces.

%%%%%%%%%%%%%%%%%%%%%%%%%%%%%%%%%%%%%%%%%%%%%%%%%%%%%%%%%%%%%%%%%%%%%%%%%%%%%%%
\section{Decomposition of $\Gamma$ in cyclic subgroups}
\label{S:CycDec}
%%%%%%%%%%%%%%%%%%%%%%%%%%%%%%%%%%%%%%%%%%%%%%%%%%%%%%%%%%%%%%%%%%%%%%%%%%%%%%%

In this section we develop some formal results in order to proceed further.
Especifically, we define the topological signature of any covariance function
that can be decomposed as
\[
X^\Gamma = \sum_{g \in \Gamma} X^g \; ,
\]
as for example, the two--point correlation function of the Newtonian
potential and the correlation matrix $\langle a_{\ell m} \, a^*_{\ell' m'}
\rangle$. Then we work out a suitable decomposition of a covering group in
their cyclic subgroups, and write down the topological signature in terms of
this decomposition. We also show here that the symmetries of a quotient space
appear transparently in the decomposition of its covering group in cyclic
subgroups. It follows that the main result of this section is the ellucidation
of how these symmetries manifest themselves in the topological signature of
CMB temperature anisotropy maps.

Let us begin by writing the obvious decomposition
\begin{equation}
\label{TopSignGen}
X^\Gamma = X^{s.c.} + X^{\widehat{\Gamma}} \; ,
\end{equation}
where $\widehat{\Gamma} = \Gamma \setminus \{id\}$. The second term in the
right hand side is the topological signature in the covariance function. The
expressions we present in the following are analogous to (\ref{TopSignGen})
and are also rather obvious.

It is convenient to introduce a notation, so natural, that has been used in
(\ref{TopSignGen}) without any previous definition. Let $S$ be any subset of
isometries of the covering space $\widetilde{M}$, then a superscript $S$ in
the covariance function means
\[
X^S = \sum_{g \in S} X^g \; .
\]

Then if $M = \widetilde{M}/\Gamma$ is a quotient space and  $\Gamma_1 \subset
\Gamma$ is any subset of the covering group, we can immediately write
$X^\Gamma = X^{\Gamma_1} + X^{\Gamma \setminus \Gamma_1}$. This expression is
nothing but the simplest generalization of Eq.
(\ref{TopSignGen}), which corresponds to the trivial case $\Gamma_1 = \{id\}$.
We get a further generalization as follows, let $\Gamma_1$ and $\Gamma_2$ be
any two subsets of the covering group $\Gamma$, such that $\Gamma_1 \cap
\Gamma_2 = \Gamma_3$, then
\begin{equation}
\label{SecDecomp}
X^\Gamma = X^{\Gamma_1} + X^{\Gamma_2} - X^{\Gamma_3} + X^{\Gamma \setminus
(\Gamma_1 \cup \Gamma_2)} \; .
\end{equation}
We can now write the formal result we are interested in. Consider the subsets
$G_1, \dots, G_n \subset \Gamma$ such that for any $i \neq j$, $G_i \cap G_j =
H$, then by induction on (\ref{SecDecomp}) we get
\begin{equation}
\label{FinDecomp}
X^\Gamma = \sum_{i=1}^n X^{G_i} - (n-1) X^H + X^{\Gamma \setminus G} \; ,
\end{equation}
where $G = \cup G_i$.

To move forward, let $G_1 = \langle g_1 \rangle$ and $G_2 = \langle g_2
\rangle$ be two cyclic subgroups of $\Gamma$, and let $\black{0} \in
\widetilde{M}$ be a lift to $\widetilde{M}$ of the position of the observer in
$M$. We will say that $g_1$ and $g_2$ are conjugate by an isometry that
``does not move the observer'' if there exists an isometry $\eta$ fixing
$\black{0}$ and such that $g_1 = \eta^{-1} g_2 \eta$. Note that, as a
consequence, we have that $d(\black{0}, g_1 \black{0}) = d(\black{0}, g_2
\black{0})$, where $d(\black{x},\black{y})$ is the distance between two points
$\black{x}$ and $\black{y}$ in $\widetilde{M}$. By extension, we will also say
that the groups $G_1$ and $G_2$ are conjugate by an isometry that does not
move the observer. In addition, we will say that $g_1$ is a minimal distance
generator of $G_1$ (with respect to the observer) if $d(\black{0},g_1
\black{0}) \leq d(\black{0},\gamma\black{0})$ for any other generator $\gamma
\in G_1$.

Now consider the isometries $g_1, \dots, g_n \in \Gamma$ that generate the
cyclic groups $G_i = \langle g_i \rangle$, and such that if $i \neq j$ then
$G_i \cap G_j = \{id\}$. By using Eq. (\ref{FinDecomp}) we immediately obtain
that the topological signature of the covariance function can be decomposed as
\[
X^{\widehat{\Gamma}} = \sum_{i=1}^n X^{\widehat{G}_i} + X^{\Gamma \setminus G}
\; .
\]
In the following we will be particularly interested in the case where the
$g_i$'s are minimal distance generators of the $G_i$'s, and the latter form a
complete set of groups mutually conjugate by isometries that do not move the
observer.

Decomposing the topological signature further along these lines, let $g_1,
\dots, g_n, h_1, \dots, h_m \in \Gamma$ be minimal distance generators of the
groups $G_i = \langle g_i \rangle$ and $H_j = \langle h_j \rangle$, and let $G
= \cup G_i$ and $H = \cup H_j$. Moreover, suppose that the $G_i$'s and the
$H_j$'s form two complete sets of groups mutually conjugate by isometries that
do not move the observer, and such that $G \cap H = \{id\}$ and $d(\black{0},
g_1 \black{0}) \leq d(\black{0}, h_1 \black{0})$. Then the topological
signature can be decomposed as
\[
X^{\widehat{\Gamma}} = \sum_{i=1}^n X^{\widehat{G}_i} + \sum_{i=1}^m
X^{\widehat{H}_i} + X^{\Gamma \setminus (G \cup H)} \; .
\]

We can proceed along these lines again and again, and obtain the following
decomposition, in cyclic subgroups, of the covering group $\Gamma$,
\begin{equation}
\label{DecCyclic}
\Gamma = \bigcup_{i=1}^\infty \bigcup_{j=1}^{k_i} \Gamma_{ij} \; ,
\end{equation}
where $g_{ij} \in \Gamma$ is a minimal distance generator of the cyclic group
$\Gamma_{ij}$, and such that
\begin{enumerate}
\item For each $i \in \mathbb{N}$, the set $\{\Gamma_{i1}, \dots,
\Gamma_{ik_i}\}$ is a complete set of groups mutually conjugate by isometries
that do not move the observer.
\item If $i \neq i'$, the sets $\cup_{j=1}^{k_i} \Gamma_{ij}$ and
$\cup_{j=1}^{k_{i'}} \Gamma_{i'j}$ have the identity as the only common
element.
\item If $i < i'$, then $d(\black{0}, g_{i1} \black{0}) \leq d(\black{0},
g_{i'1} \black{0})$.
\end{enumerate}
Then the topological signature of the covariance function can be written as
\begin{equation}
\label{DecTSCyclic}
X^{\widehat{\Gamma}} = \sum_{i=1}^\infty \sum_{j=1}^{k_i}
X^{\widehat{\Gamma}_{ij}} \; .
\end{equation}
Thus, to compute the topological signature of the covariance function for any
multiply connected manifold, it is enough to know how to compute it for
manifolds whose covering groups are cyclic groups. It is now obvious that this
decomposition will be particularly useful for calculating the correlation
matrix of the $a_{\ell m}$'s for any compact manifold once we know how to
calculate it for cyclic flat (twisted cylinders), spherical (lens spaces), and
hyperbolic manifolds.

Let us now show that the decomposition (\ref{DecCyclic}) describes
transparently the symmetries of the quotient manifold $M =
\widetilde{M}/\Gamma$. Actually, this decomposition contains the symmetries of
the Dirichlet fundamental polyhedron of $M$ centered at the observer's
position $\black{0} \in \widetilde{M}$, which is what one expects to
reconstruct with cosmological observations. Recall that the Dirichlet
fundamental polyhedron centered at $\black{0} \in \widetilde{M}$ is the set
$\mathcal{D}_{\black{0}} \subset \widetilde{M}$ defined by (see
\cite{Beardon})
\[
\mathcal{D}_{\black{0}} = \{ \black{x} \in \widetilde{M} : d(\black{0},
\black{x}) \leq d(g\black{0},\black{x}) \; \mbox{ for any } \; g \in \Gamma \}
\; .
\]
The first thing to be noted is that, although the whole covering group enters
in this definition, it turns out that, in order to effectively construct the
Dirichlet polyhedron, we only need the minimal distance generators (and maybe
the first few positive powers) of the first few cyclic groups $\Gamma_{ij}$
and their inverses.

In fact, for each $g \in \Gamma$ consider the semi--space
\[
H_g = \{ \black{x} \in \widetilde{M} \; : \; d(\black{0},\black{x}) \leq
d(g\black{0},\black{x}) \} \; .
\]
Then it is obvious that the Dirichlet polyhedron is the intersection of all of
these semi--spaces. However, there is a high redundancy here, since for a
sufficiently large positive power $n$, we may have
\[
\bigcap_{k=1}^{n-1} H_{g_{ij}^k} \subset H_{g_{ij}^n} \; ,
\]
and so this and further powers of $g_{ij}$ do not effectively contribute to
the polyhedron $\mathcal{D}_{\black{0}}$. Additionally, if some $H_g$
effectively contributes to the polyhedron, so does $H_{g^{-1}}$, thus the same
argument holds for the inverses of the minimal distance generators. Moreover,
note that due to condition 3. above, for a sufficiently large $i$, it may be
the case that the semi--spaces $H_{g_{ij}}$ do not contribute effectively to
the polyhedron.

The faces of the Dirichlet polyhedron are subsets of the boundary planes of
the semi--spaces effectively contributing to it. In fact, for each $H_g$
effectively contributing, the corresponding face is orthogonal to the geodesic
joining $\black{0}$ and $g\black{0}$, and cuts it at its middle point. It
follows that the decomposition (\ref{DecCyclic}) describes the symmetries of
the Dirichlet fundamental polyhedron of $M$ centered at the observer.

%%%%%%%%%%%%%%%%%%%%%%%%%%%%%%%%%%%%%%%%%%%%%%%%%%%%%%%%%%%%%%%%%%%%%%%%%%%%%%%
\section{Flat homogeneous manifolds}
\label{S:tori}
%%%%%%%%%%%%%%%%%%%%%%%%%%%%%%%%%%%%%%%%%%%%%%%%%%%%%%%%%%%%%%%%%%%%%%%%%%%%%%%

We have seen in the previous section that, to compute the topological
signature of CMB temperature anisotropies in a given manifold, we just need to
know how to compute it for the cyclic manifolds that cover it maximally. In
the flat orientable case the cyclic manifolds are twisted cylinders, i.e.,
manifolds with covering group generated by a screw motion. We will now focus
on the simplest case, the flat homogeneous manifolds, which are generated by
translations only.

The flat homogeneous manifolds are 3--tori or $T^3$ manifolds (generated by
three linearly independent translations), chimneys or $T^2$ manifolds
(generated by two linearly independent translations), and cylinders or $T^1$
manifolds (generated by one translation). Thus, we first compute the
correlation matrix $\langle a_{\ell m} \, a^*_{\ell' m'} \rangle^{\Gamma}$ for
cylinders, and then show how the decomposition (\ref{DecTSCyclic}) is used to
compute the topological signature in this matrix for two-- and
three--dimensional tori. We also show that the computation of the angular
power spectrum in tori is greatly simplified by this decomposition.

%%%%%%%%%%%%%%%%%%%%%%%%%%%%%%%%%%%%%%%%%%%%%%%%%%%%%%%%%%%%%%%%%%%%%%%%%%%%%%%
\subsection{The cylinder}
\label{Ss:Cylinder}
%%%%%%%%%%%%%%%%%%%%%%%%%%%%%%%%%%%%%%%%%%%%%%%%%%%%%%%%%%%%%%%%%%%%%%%%%%%%%%%

Let us consider a cylinder orthogonal to the $z$--direction, that is with
covering group generated by the translation $g = (I, \black{a})$, with
$\black{a} = L \base{z}$, where distances are measured in units of the radius
of the last scattering surface $R_{LSS}$. This choice of the coordinate system
is very convenient since here the cylinder appears invariant under (i)
arbitrary rotations around the $z$--axis, (ii) the parity transformation, and
(iii) the reflection on the $y=0$ plane. Thus, according to Sec.
\ref{Ss:Symmetry}, we will end with a real correlation matrix with no
$m$--dependent correlations and the multipoles $\ell$ and $\ell'$ correlated
only when both are even or odd.

It is convenient to recall here that our cylinder has injectivity radius equal
to $L/2$, thus cylinders with $L<2$ are ``small'' and might present
topological copies of discrete sources and/or circles in the sky. On the other
hand, ``large'' cylinders, i.e., those with $L>2$, have undetectable topology
with the methods currently available \cite{Detect}.

The covering group of the cylinder is labeled by the integers as $g^n = (I,n
\black{a})$, with $n \in \mathbb{Z}$. We then have from (\ref{DefTopSign})
that all the topological information is encoded in
\[
\Upsilon^{\Gamma}_{\ell m} (\black{k}) = \sum_{n \in \mathbb{Z}} e^{-i nk_zL}
\, Y_{\ell m}(\black{n}_{\black{k}}) \; ,
\]
and thus the correlation matrix of the $a_{\ell m}$'s for the cylinder is
simply
\begin{widetext}
\begin{equation}
\label{CorrMatCylGen}
\langle a_{\ell m} \, a^*_{\ell' m'} \rangle^{\Gamma} = (4 \pi)^2 \, i^{\ell - \ell'} \int \!
\frac{d^3k}{k^3} \, \Psi_{\ell \ell'}(k) \left( \sum_{n \in \mathbb{Z}} e^{-i nk_zL}
\right) Y_{\ell' m'}(\black{n}_{\black{k}}) \, Y_{\ell m}^*(\black{n}_{\black{k}})\; .
\end{equation}
\end{widetext}

To reduce this integral we may use any of the following two identities, either
\begin{equation}
\label{FirstId}
\sum_{n \in \mathbb{Z}} e^{-i nk_zL} = 2\pi \sum_{p \in \mathbb{Z}} \delta(k_zL
- 2\pi p) \; ,
\end{equation}
or
\begin{equation}
\label{SecondId}
\sum_{n \in \mathbb{Z}} e^{-i nk_zL} = 1 + 2\sum_{n=1}^\infty \cos(nk_zL) \; .
\end{equation}
The first identity is obvious since the left hand side is the Fourier
expansion of the right hand side. This option yields a formula of the kind
obtained in Appendix \ref{S:OldRes}, which expresses the correlation matrix in
terms of the eigenvalues of the Laplacian operator on the cylinder. The second
one still uses a parametrization in terms of the covering group, and thus can
be used to isolate the topological signature.

\begin{figure*}[t]
\setlength{\unitlength}{1cm}
\begin{picture}(10,4.8)
\scalebox{0.35}{
\includegraphics[8cm,5cm][10cm,8cm]{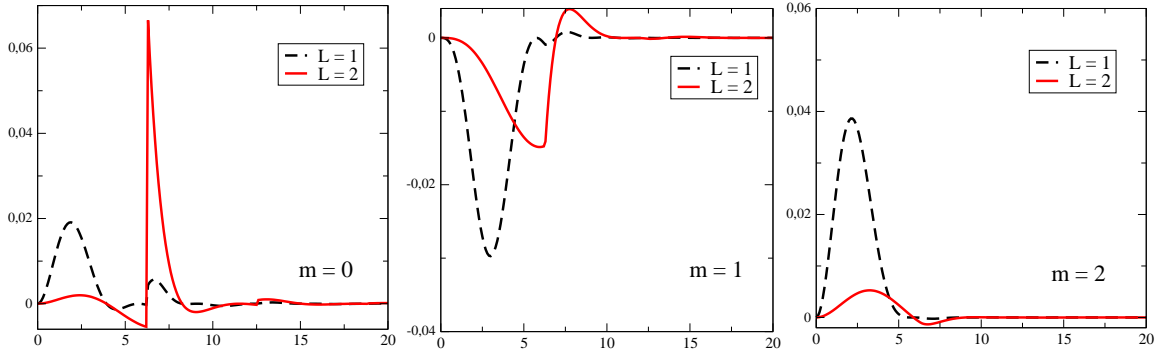}
}
\end{picture}
\caption{\label{fig:fm22} \small Shapes of the integrands inside integrals of
the type in (\ref{TopSignCyl}) for $\ell = \ell' = 2$, and compactification
scales $L=1$ and $L=2$ in units of the radius of the last scatering surface.}
\end{figure*}

In fact, using (\ref{SecondId}) to evaluate (\ref{CorrMatCylGen}), and
integrating in spherical coordinates, we get
\begin{equation}
\label{CorrMatCyl}
\langle a_{\ell m} \, a^*_{\ell' m'} \rangle^{\Gamma} = \langle a_{\ell m} \,
a^*_{\ell' m'} \rangle^{s.c.} + \langle a_{\ell m} \, a^*_{\ell' m'}
\rangle^{\widehat{\Gamma}} \; ,
\end{equation}
where the simply connected part is as usual
\begin{equation}
\label{CorrMatSC}
\langle a_{\ell m} \, a^*_{\ell' m'} \rangle^{s.c.} = C_\ell^{s.c.}
\delta_{\ell \ell'} \delta_{mm'} \; ,
\end{equation}
with the (simply connected) angular power spectrum given by
\begin{equation}
\label{PowSpecSC}
C_\ell^{s.c.} = (4\pi)^2 \int_0^\infty \frac{dx}{x} \, \Psi_{\ell \ell}(x)
\; ,
\end{equation}
and the topological signature for the correlation matrix is given by
\begin{eqnarray}
\label{TopSignCyl}
\langle a_{\ell m} \, a^*_{\ell' m'} \rangle^{\widehat{\Gamma}} & = & (4\pi)^2
\, i^{\ell - \ell'} \delta_{\ell \ell'}^{\mbox{\tiny mod(2)}} \, \delta_{mm'}
\times \\
& & \hspace{1cm} \int_0^\infty \frac{dx}{x} \, \Psi_{\ell \ell'} \left(
\frac{x}{L} \right) F_{\ell \ell'}^m(x) \; , \nonumber
\end{eqnarray}
with
\begin{equation}
\label{CylFllmInit}
F_{\ell \ell'}^m(x) = 2 \sum_{n=1}^\infty \int_{-1}^1 \! dy \, \cos(nxy)
\mathcal{P}_\ell^m(y) \mathcal{P}_{\ell'}^m(y) \; ,
\end{equation}
where $\mathcal{P}_\ell^m(x)$ is the normalized associated Legendre function
(see Appendix \ref{Ap:SphHarm}). As expected, we have ended up with a real
correlation matrix with factors $\delta_{\ell \ell'}^{\mbox{\tiny mod(2)}}$
and $\delta_{m m'}$.

After evaluating the series in (\ref{CylFllmInit}), it turns out that $F_{\ell
\ell'}^m(x)$ is a piecewise continuous function. In fact, in each interval
$[2\pi q, 2\pi (q+1)]$, it is a polynomial of degree $(\ell + \ell' + 1)$ in
$\pi/x$. Indeed, the final result is
\begin{eqnarray}
\label{CylFllmFin}
F_{\ell \ell'}^m(x) & = & \sum_{q \in \mathbb{Z}} \mathcal{F}_{\ell
\ell'}^m(x,q) \, \Theta(x - 2\pi q)  \times \\
& & \hspace{2.5cm} \Theta(2\pi(q+1) - x) \; , \nonumber
\end{eqnarray}
where $\Theta(x)$ is the Heaviside step function, and the form of the
polynomial $\mathcal{F}_{\ell \ell'}^m(x,q)$ in the $q$--th interval of length
$2\pi$ is
\begin{eqnarray}
\label{PolyqCylFllmFin}
\mathcal{F}_{\ell \ell'}^m(x,q) & = & 4 \sum_{k=0}^{\frac{\ell + \ell'}{2}}
(-1)^k \mathcal{P}_{\ell \ell' m}^{(2k)}(0) \times \\
& & \hspace{1cm} g_{2k+1}(q) \, \left( \frac{\pi}{x} \right)^{2k+1} -
\delta_{\ell \ell'} \; . \nonumber
\end{eqnarray}
Here $g_k(q)$ are polynomials of degree $k$ in $q$, and $\mathcal{P}_{\ell
\ell' m}^{(k)}(0)$ is the $k$--th derivative of the polynomial
\begin{equation}
\label{PllmDef}
\mathcal{P}_{\ell \ell' m}(x) = \mathcal{P}_\ell^m(x) \mathcal{P}_{\ell'}^m(x)
\end{equation}
evaluated at the origin. In Appendix \ref{Ap:Clausen} we present recurrence
relations for the polynomials $g_k(q)$, and all the technical steps that take
us from (\ref{CylFllmInit}) to (\ref{CylFllmFin}) can be found in Appendix
\ref{Ap:EvalFllm}.

\begin{figure*}[t]
\setlength{\unitlength}{1cm}
\begin{picture}(8,7.5)
\scalebox{0.5}{
\includegraphics[7cm,2.5cm][10cm,10cm]{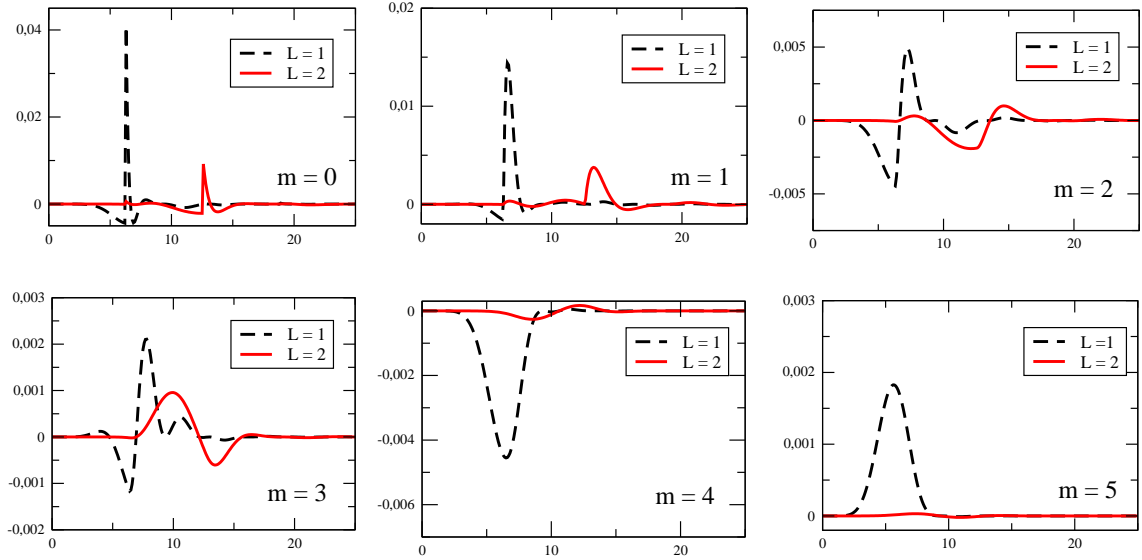}
}
\end{picture}
\caption{\label{fig:fm55} \small Shapes of the integrands inside integrals of
the type in (\ref{TopSignCyl}) for $\ell = \ell' = 5$, and compactification
scales $L=1$ and $L=2$ in units of the radius of the last scatering surface.}
\end{figure*}

The integrals appearing in the topological signature (\ref{TopSignCyl}) can be
easily evaluated since Eqs. (\ref{CylFllmFin}) and (\ref{PolyqCylFllmFin})
allow an exact and very fast computation of the function $F_{\ell
\ell'}^m(x)$, and the integrands decay very fast, as illustrated in
figs.\ref{fig:fm22} and \ref{fig:fm55}. In this figures we have adopted, for
simplicity, a scale invariant Einstein--de Sitter model, thus
\begin{equation}
\label{PsiEdS}
\Psi_{\ell \ell'}(x) \propto j_\ell (x)j_{\ell'}(x) \; .
\end{equation}
The nice behavior of the integrands in (\ref{TopSignCyl}) is not a consequence
of this particular choice of $\Psi_{\ell \ell'}(x)$. Actually, the integrand
in (\ref{TopSignCyl}) always decays very fast because $\Psi_{\ell \ell'}(x)$
and $F_{\ell \ell'}^m(x)$ are both decaying functions, thus the evaluation of
the topological signature for the cylinder is always very efficient.

The computation of the topological signature of the power spectrum reduces to
a simple integral. In fact we obtain
\[
C_\ell^{\widehat{\Gamma}} = (4\pi)^2 \int_0^\infty \frac{dx}{x} \, \Psi_{\ell
\ell} \left( \frac{x}{L} \right) f_\ell(x) \; ,
\]
with
\[
f_\ell(x) = \frac{1}{2\ell + 1} \sum_{m=-l}^\ell F_{\ell \ell}^m(x) \; .
\]
Using (\ref{CylFllmInit}) to perform this sum, the Addition Theorem for
Spherical Harmonics (see Appendix \ref{Ap:SphHarm}) yields immediately
\begin{equation}
\label{TopSignClFin}
C_\ell^{\widehat{\Gamma}} = 2(4\pi)^2 \int_0^\infty \frac{dx}{x^2} \,
\Psi_{\ell \ell} \left( \frac{x}{L} \right) \varphi_1(x) \; ,
\end{equation}
where $\varphi_1(x)$ is the first Clausen $\varphi$--function given in
Appendix \ref{Ap:Clausen}.

In fig.\ref{fig:senhalCl_L}a we show the low $\ell$--modes of the topological
signature of the angular power spectrum of a cylinder, normalized w.r.t.
$C_{\ell}^{s.c.}$, as a function of its size $L$. We can see that the
topological signature is typically much smaller than the cosmic variance, even
for small cylinders which have already been discarded observationally as
candidates for the shape of our Universe because of the lack of antipodal
matched circles in WMAP data \cite{Aligne1, CSSK04}. Thus it is apparent that
the angular power spectrum is not a good indicator to look for topology in
this case.

The correlation matrix given by (\ref{CorrMatCyl})--(\ref{PolyqCylFllmFin})
corresponds to a cylinder for which the direction of compactification is
parallel to the $z$--axis. The correlation matrix corresponding to a cylinder
with a different orientation can be easily obtained from the previous one by
simply rotating the celestial sphere. Thus, parametrizing the rotations with
Euler angles, if $R(\alpha, \beta, \gamma) \in SO(3)$ takes the $z$--axis to
the direction of compactification of the cylinder, the topological signature
of the corresponding correlation matrix can be computed using the expressions
(\ref{a_lmRot})--(\ref{Wigner-Euler}) of Appendix \ref{Ap:SphHarm} yielding
\begin{eqnarray}
\label{CorrMatCylRot}
\langle a_{\ell m} \, a^*_{\ell' m'} \rangle^{\widehat{\Gamma}}_R & = &
e^{i(m'-m) \alpha} \times \\
& & \hspace{0.5cm} \sum_{m_1}  d_{mm_1}^\ell(\beta) \,
d_{m'm_1}^{\ell'}(\beta) \times \nonumber \\
& & \hspace{2cm} \langle a_{\ell m_1} \, a^*_{\ell' m_1}
\rangle^{\widehat{\Gamma}} \; , \nonumber
\end{eqnarray}
since $\langle a_{\ell m} \, a^*_{\ell' m'} \rangle^{s.c.}$ is rotationally
invariant. Moreover, the $\gamma$ angle does not appear in this expression
since $R_z(\gamma)$ in (\ref{Euler}) does not move the $z$--axis, and $\langle
a_{\ell m} \, a^*_{\ell' m'} \rangle^{\Gamma}$ is invariant under such
rotations.

It should be noted here that, no matter its orientation, the cylinder is
always invariant under parity, thus its correlation matrix will always
conserve the factor $\delta_{\ell \ell'}^{\mbox{\tiny mod(2)}}$. On the other
hand, the correlation matrix will remain real as far as we perform rotations
with $\alpha = 0$, since in this case we do not rotate the cylinder around the
$z$--axis, and thus it remains invariant under reflections on the plane $y=0$.
However, any rotation of the cylinder (other than one with $\beta = \pi$)
makes it non--invariant under azimuthal rotations, thus the correlation matrix
of an arbitrarily oriented cylinder has $m$--dependent correlations. All these
features can be seen explicitly in (\ref{CorrMatCylRot}).

%%%%%%%%%%%%%%%%%%%%%%%%%%%%%%%%%%%%%%%%%%%%%%%%%%%%%%%%%%%%%%%%%%%%%%%%%%%%%%%
\subsection{tori}
\label{Ss:tori}
%%%%%%%%%%%%%%%%%%%%%%%%%%%%%%%%%%%%%%%%%%%%%%%%%%%%%%%%%%%%%%%%%%%%%%%%%%%%%%%

In order to calculate the correlation matrix of the $a_{\ell m}$'s for a two--
or a three--torus we use the decomposition (\ref{DecCyclic}) of its covering
group in cyclic subgroups. Let $\Gamma_{ij} = \langle g_{ij} \rangle$ be the
covering group of the cylinder generated by the element $g_{ij} \in \Gamma$,
and let us write $L_i = d(\black{0}, g_{ij} \black{0})$, $g_i = (I, L_i
\base{z})$, and $\Gamma_i = \langle g_i \rangle$. In the Euclidean case, the
orientation preserving isometries that do not move the observer are rotations,
thus let $R_{ij} \in SO(3)$ be the rotation taking $\base{z}$ to the unit
vector along $g_{ij} \black{0}$.

\begin{figure*}[t]
\setlength{\unitlength}{1cm}
\begin{picture}(6,5.3)
\scalebox{0.4}{
\includegraphics[10cm,5.2cm][20cm,13cm]{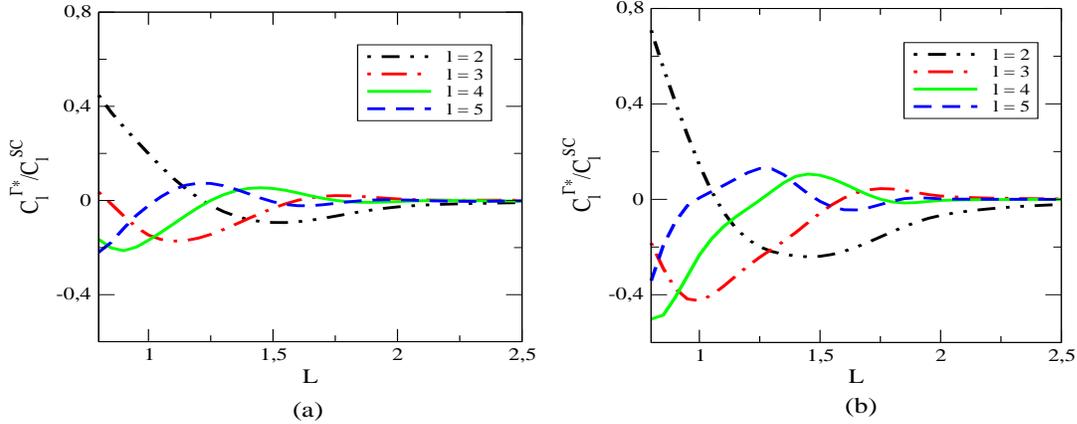}
}
\end{picture}
\caption{\label{fig:senhalCl_L} \small Topological signature of the power
spectrum of (a) a cylinder, and (b) a chimney with square base, for the first
four $\ell$--modes, normalized w.r.t. $C_{\ell}^{s.c.}$, and as a function of
the scale of compactification $L$. Note that for each multipole one can have a
suppression or an excess of power depending on the value of $L$. For
typographical reasons, here we write $\Gamma^*$ instead of
$\widehat{\Gamma}$.}
\end{figure*}

Using the decomposition (\ref{DecTSCyclic}), we easily write the topological
signature for the torus as a superposition of topological signatures of
rotated cylinders. In fact,
\begin{equation}
\label{TopSignTorusDef}
\langle a_{\ell m} \, a^*_{\ell' m'} \rangle^{\widehat{\Gamma}} =
\sum_{i=1}^\infty \sum_{j=1}^{k_i} \langle a_{\ell m} \, a^*_{\ell' m'}
\rangle^{\widehat{\Gamma}_i}_{R_{ij}} \; ,
\end{equation}
where the correlation matrices of the rotated cylinders are written in terms
of the Wigner $D$--functions and Euler angles, according to
(\ref{CorrMatCylRot}), as
\begin{eqnarray}
\label{TopSignTorusAux}
\langle a_{\ell m} \, a^*_{\ell' m'} \rangle^{\widehat{\Gamma}_i}_{R_{ij}}
& = & e^{i(m'-m) \alpha_{ij}} \times \\
& & \sum_{m_1} d_{mm_1}^\ell(\beta_{ij}) \, d_{m'm_1}^{\ell'}(\beta_{ij})
\times \nonumber \\
& & \hspace{2cm} \langle a_{\ell m_1} \, a^*_{\ell' m_1}
\rangle^{\widehat{\Gamma}_i} \; , \nonumber
\end{eqnarray}
and $(\beta_{ij}, \alpha_{ij})$ are the angular spherical coordinates of the
vector $g_{ij} \black{0}$, and $k_i$ is the number of cylinders of size $L_i$.
Since any group of translations is invariant under parity, from Sec.
\ref{Ss:Symmetry} we know that the correlation matrix for a homogeneous flat
manifold has always the factor $\delta_{\ell \ell'}^{\mbox{\tiny mod(2)}}$,
and this is evident from (\ref{TopSignTorusDef}), since it is just a sum of
correlation matrices of cylinders.

The power spectrum is rotationally invariant, thus from
(\ref{TopSignTorusDef}) one can easily write down the expression for the
topological signature of the power spectrum of the torus as a superposition of
topological signatures of power spectra of cylinders,
\begin{equation}
\label{ClTorus}
C_\ell^{\widehat{\Gamma}} = \sum_{i=1}^\infty k_i \,
C_\ell^{\widehat{\Gamma}_i} \; .
\end{equation}

Let us consider a chimney with square base for the sake of illustration. It is
convenient to orient the chimney so that its covering group consists of
translations in the horizontal plane. We take as generators of the covering
group the translations $g_1 = (I, \black{a})$ and $g_2 = (I, \black{b})$, with
$\black{a} = L \base{x}$ and $\black{b} = L \base{y}$.

It is more convenient to reparametrize the cyclic decomposition as follows.
Parametrize each cyclic subgroup by a pair of integer numbers $(p,q)$ as
$G_{pq} = \langle g_2^q g_1^p \rangle$. Clearly, if the greatest common
divisor of $(p,q)$ is $r$, then
\[
G_{pq} < G_{\frac{p}{r} \frac{q}{r}} \; ,
\]
where `$<$' means `subgroup of'. Thus we must restrict the labels to pairs
$(p,q)$ of coprime numbers.

The only exceptions are when (i) $p=\pm 1$ and $q=0$ and viceversa, and (ii)
when $p=\pm 1$ and $q=\pm 1$. Thus the first two complete sets of cyclic
subgroups conjugate by a rotation are $\{ G_{1,0} , G_{0,1} \}$ and $\{
G_{1,1} , G_{-1,1} \}$. In both cases the conjugation is performed by a
rotation of $\pi/2$ around the $z$--axis. The compactification lengths of the
corresponding cylinders are $L_{1,0} = L_{0,1} = L$ and $L_{1,1} = L_{-1,1} =
\sqrt{2} L$ respectively. The Euler angles $(\beta, \alpha)$ to rotate the
corresponding cylinders from the $z$--axis to their orientation in the
chimney, according to (\ref{TopSignTorusAux}), are $\beta = \pi/2$ in all
cases, and $\alpha_{1,0} = 0$, $\alpha_{0,1} = \pi/2$, $\alpha_{1,1} = \pi/4$
and $\alpha_{-1,1} = 3\pi/4$, respectively.

To write the remaining complete sets of cyclic subgroups conjugate by a
rotation let us define, for a pair of coprime natural numbers $(p,q)$, with
$p>q\geq1$, the groups
\begin{eqnarray*}
G_{pq}^{(1)} = G_{pq} = \langle g_2^q g_1^p \rangle & , & G_{pq}^{(3)} =
G_{-q,p} = \langle g_2^p g_1^{-q} \rangle \; , \\
G_{pq}^{(2)} = G_{qp} = \langle g_2^p g_1^q \rangle & , & G_{pq}^{(4)} =
G_{-p,q} = \langle g_2^q g_1^{-p} \rangle \; .
\end{eqnarray*}
The compactification lengths are all equal to $L_{pq} = \sqrt{p^2+q^2}L$, and
the Euler angles $(\beta, \alpha)$ to rotate the corresponding cylinders from
the $z$--axis to their orientation in the chimney, according to
(\ref{TopSignTorusAux}), are $\beta = \pi/2$ in all cases, and
\begin{eqnarray*}
\alpha_{pq}^{(1)} = \arctan \frac{q}{p} & , & \alpha_{pq}^{(3)} =
\frac{\pi}{2} + \alpha_{pq}^{(1)} \; , \\
\alpha_{pq}^{(2)} = \frac{\pi}{2} - \alpha_{pq}^{(1)} & , & \alpha_{pq}^{(4)}
= \pi - \alpha_{pq}^{(1)} \; ,
\end{eqnarray*}
respectively.

Let us denote by $\Gamma_{pq}$ the covering group of the cylinder with
compactification scale $L_{pq}$ and oriented along the $z$--axis. Then,
putting all this together, using (\ref{TopSignTorusDef}) and
(\ref{TopSignTorusAux}), and taking into account the invariance properties
derived in Sec. \ref{Ss:Symmetry}, the topological signature of the chimney
with square base is
\begin{eqnarray}
\label{TopSignChim}
\langle a_{\ell m} \, a^*_{\ell' m'} \rangle^{\widehat{\Gamma}} & = &
\delta_{mm'}^{\mbox{\tiny mod(4)}} \sum_{m_1} d_{mm_1}^\ell(\pi/2) \times \\
& & \hspace{1.5cm} d_{m'm_1}^{\ell'}(\pi/2) \, \mathcal{W}_{\ell  \ell'
m_1}^{m' - m} \; , \nonumber
\end{eqnarray}
with
\begin{widetext}
\begin{equation}
\label{WTopSignChim}
\mathcal{W}_{\ell \ell' m_1}^m = 2 \left( \langle a_{\ell m_1} \, a^*_{\ell'
m_1} \rangle^{\widehat{\Gamma}_{1,0}} + (-1)^{m/4} \langle a_{\ell m_1} \,
a^*_{\ell' m_1} \rangle^{\widehat{\Gamma}_{1,1}} \right) + 4 \sum_{(p,q)}
\cos m \alpha_{pq}^{(1)} \, \langle a_{\ell m_1} \, a^*_{\ell' m_1}
\rangle^{\widehat{\Gamma}_{pq}} \; ,
\end{equation}
\end{widetext}
where the sum in $(p,q)$ is evaluated only for pairs of coprime natural
numbers $(p,q)$ such that $p>q\geq1$.

\begin{figure*}[t]
\setlength{\unitlength}{1cm}
\begin{picture}(8,10.1)
\scalebox{0.6}{
\includegraphics[3.5cm,0.6cm][10cm,5cm]{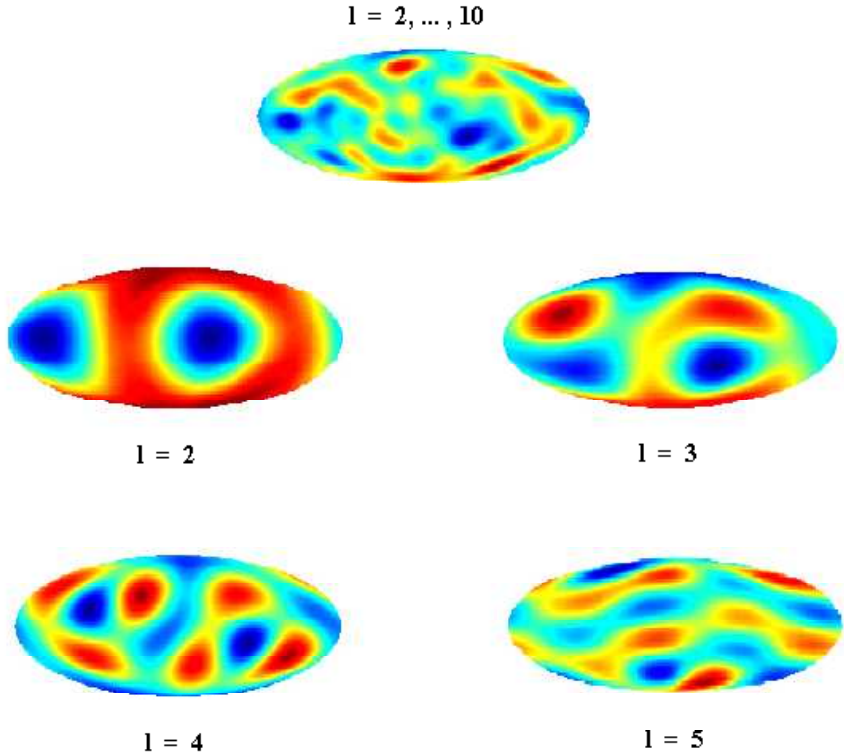}
}
\end{picture}
\caption{\label{fig:mapas} \small Simulated CMB temperature anisotropy map at
low resolution ($2 \leq \ell \leq 10$) for a universe with a $T^1$ topology
and size $L=2$. Also shown are the lowest multipoles, all of which present
clear alignments near the polar direction.}
\end{figure*}

The topological signature of the power spectrum of the chimney with square
base is simply
\begin{equation}
\label{PowSpecChim}
C_\ell^{\widehat{\Gamma}} = 2 \left( C_\ell^{\widehat{\Gamma}_{1,0}} +
C_\ell^{\widehat{\Gamma}_{1,1}} \right) + 4 \sum_{(p,q)}
C_\ell^{\widehat{\Gamma}_{pq}} \; .
\end{equation}
Since the topological signature of the power spectrum of a cylinder converges
quickly to zero as a function of the compactification scale (see
fig.\ref{fig:senhalCl_L}a), it follows that the sum in (\ref{PowSpecChim}) also
converges quickly.

Moreover, the topological signature of the power spectrum of the chimney is
larger than that of the cylinder. This is so because the $\ell$-th mode of the
topological signature of the angular power spectrum of the cylinder oscillates
very slowly. Thus from (\ref{PowSpecChim}) this signature is slightly higher
in the chimney, as can be seen in fig.\ref{fig:senhalCl_L}b. Actually, this is
a general result that holds for manifolds whose covering groups are not
cyclic.

%%%%%%%%%%%%%%%%%%%%%%%%%%%%%%%%%%%%%%%%%%%%%%%%%%%%%%%%%%%%%%%%%%%%%%%%%%%%%%%
\section{Patterns of alignment}
\label{S:TopSign}
%%%%%%%%%%%%%%%%%%%%%%%%%%%%%%%%%%%%%%%%%%%%%%%%%%%%%%%%%%%%%%%%%%%%%%%%%%%%%%%

The nondiagonal character of the topological signature of the correlation
matrix of the $a_{\ell m}$'s in multiply connected universes and their
$m$--dependence are manifestations of their globally anisotropic nature. They
manifest themselves in statistically anisotropic temperature maps, i.e.,
realizations of random temperature fluctuations for which mean values of
functions of the temperature over ensembles of universes depend on the
orientation \cite{StatAnis}.

In this section we analyze an expected consequence of the topology of space on
the temperature anisotropies of the CMB that has not received the deserved
attention up to the present, namely the existence of preferred directions in
space. We show that the decomposition of the topological signature of the
correlation matrix of the $a_{\ell m}$'s in a universe with a complex
topology, in signatures corresponding to cyclic topologies, demands the
existence of ``patterns of alignments'' along these directions. For the sake
of simplicity we consider the Einstein--de Sitter model, thus from now on we
will take (\ref{PsiEdS}) to perform all our calculations.

We want to call attention to the existence of alignments of the low
$\ell$--modes of the CMB temperature maps in multiply connected universes.
Indeed, in fig.\ref{fig:mapas} we show a low resolution temperature map
simulation for a cylinder with $L=2$ (in units of $R_{LSS}$), together with
the maps corresponding to the first four $\ell$--modes. One can see that these
$\ell$--maps present alignments along the $z$--direction, which in this case
is the unique direction of compactification of space.

\begin{figure*}[t]
\setlength{\unitlength}{1cm}
\begin{picture}(8,8.4)
\scalebox{0.5}{
\includegraphics[5.5cm,2.5cm][9cm,8cm]{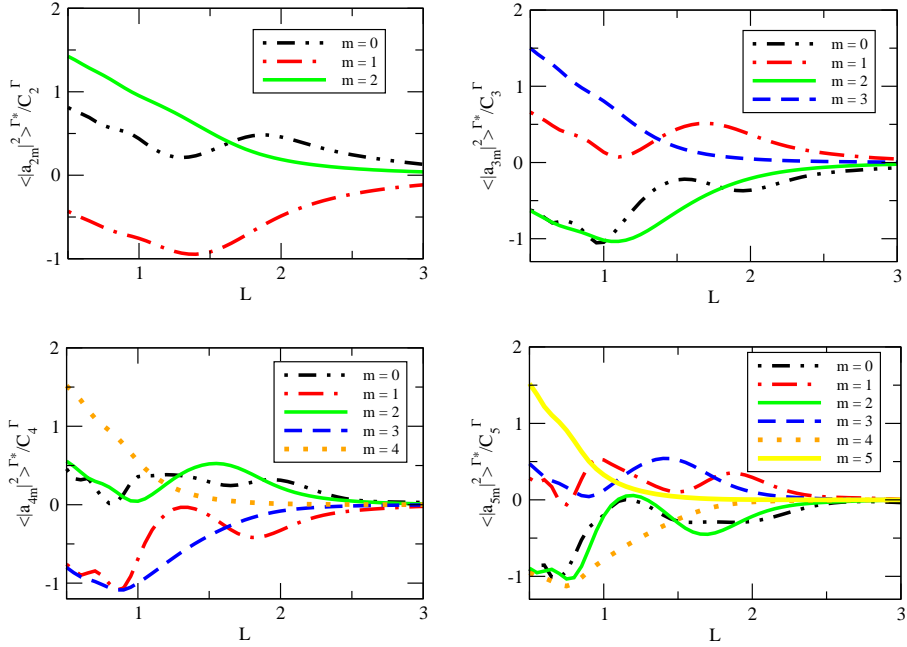}
}
\end{picture}
\caption{\label{fig:alm} \small Plots of the topological signatures of the
squares $\langle |a_{\ell m}|^2 \rangle^{\widehat{\Gamma}}$, normalized w.r.t.
$C_{\ell}^\Gamma$, for low multipoles ($2 \leq \ell \leq 5$), in a universe
with cylindrical topology aligned with the polar axis, and as a function of
the scale of compactification $L$. For typographical reasons, here we write
$\Gamma^*$ instead of $\widehat{\Gamma}$.}
\end{figure*}

Similar alignments as those present in our simulations have been reported as
being observed in WMAP data, and have been attributed to a possible nontrivial
topology of space with the shape of a cylinder \cite{Aligne1}. These models
have been quickly abandoned due to the lack of circles in the sky which should
be present if the Universe were small \cite{Aligne1, CSSK04}. However our
simulations show that even in universes slightly larger, and so not presenting
such circles, these alignments should still be observable. Thus whether these
observed alignments are a consequence of a nontrivial shape of our Universe is
still an open question \cite{LocShape}.

We will show here that if our Universe had a nontrivial topology, its CMB
temperature map will present characteristic patterns of alignment, even if its
size is somewhat larger than the observable universe. Moreover, from the
observed patterns of alignment, we might be able to reconstruct the shape of
space.

In fig.\ref{fig:alm} we show the topological signature of the low $\ell$--modes
of the diagonal part of the correlation matrix of the $a_{\ell m}$'s,
normalized w.r.t. $C_{\ell}^\Gamma$, for a cylinder oriented along the polar
axis, as a function of the size $L$ of compactification. It is apparent that,
for a given $\ell$--mode, there are multipole coefficients for which their
expected values are above the mean (the angular power spectrum), and others
for which these expected values are below it. This is the reason why the low
$\ell$--modes in a cylinder are aligned. Actually, the expectation values
$\langle |a_{\ell m}|^2 \rangle$ are all equal to $C_\ell$ only in the simply
connected case, thus the dispersion around the mean
\begin{equation}
\label{Dispalm}
\sigma_\ell = \sqrt{\frac{1}{2 \ell + 1} \sum_m \left( \langle |a_{\ell m}|^2
\rangle - C_\ell \right)^2}
\end{equation}
is null. However, in a multiply connected universe, this dispersion is non
zero, and in a particular map, it adds to the cosmic variance. Thus it seems
natural to propose the dispersion of the squares $|a_{\ell m}|^2$ around their
mean value as a measure of these alignments in a map.

\begin{figure}[b]
\setlength{\unitlength}{1cm}
\begin{picture}(5,5.1)
\scalebox{0.3}{
\includegraphics[5cm,2.3cm][8cm,9cm]{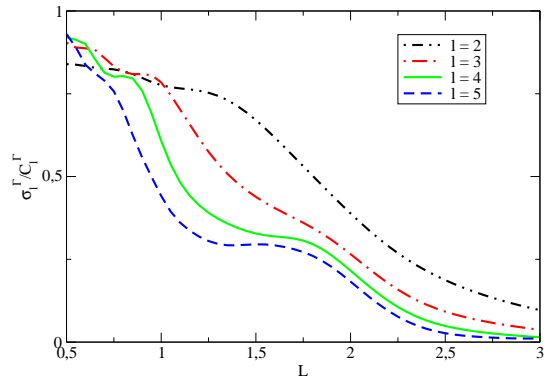}
}
\end{picture}
\caption{\label{fig:dispcyl} \small Dispersion of the $\langle |a_{\ell m}|^2
\rangle^\Gamma$, normalized w.r.t. $C_\ell^\Gamma$, for low multipoles ($2 \leq
\ell \leq 5$), in a universe with cylindrical topology aligned with the polar
axis, and as a function of the scale of compactification $L$.}
\end{figure}

In fig.\ref{fig:dispcyl} we show a plot of the dispersion (\ref{Dispalm}),
normalized w.r.t. $C_\ell^\Gamma$, for a cylinder oriented along the polar
axis, as a function of $L$, for low multipoles. Note that even for a large
cylinder ($L \approx 2$) the dispersion is larger than $15 \%$ of the power
for multipoles up to $\ell = 5$. Indeed, on these scales the dispersion is of
the order of the cosmic variance, and thus might be detectable.

\begin{figure*}[t]
\setlength{\unitlength}{1cm}
\begin{picture}(8,8.5)
\scalebox{0.5}{
\includegraphics[5cm,2.4cm][12cm,12cm]{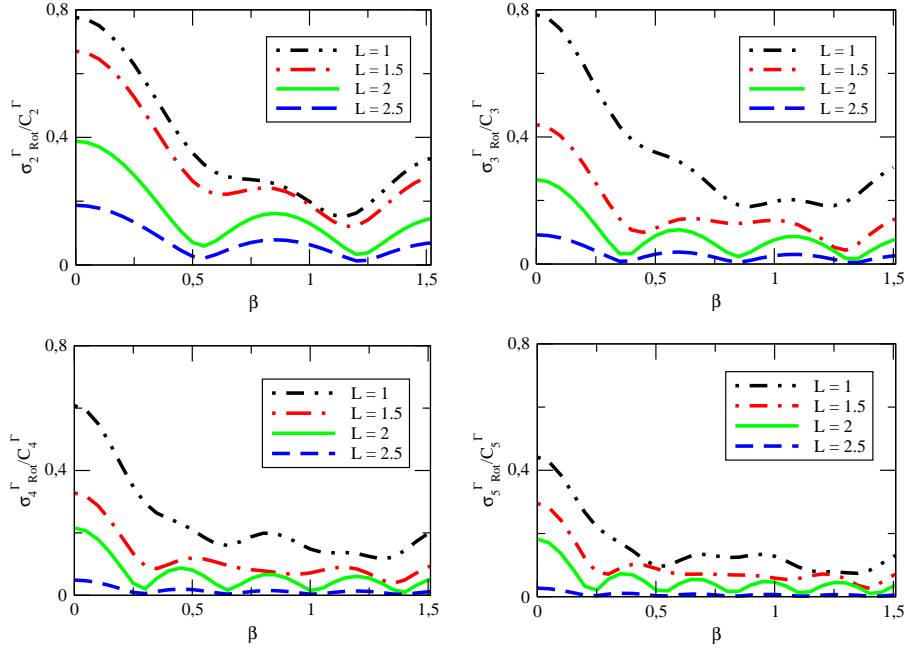}
}
\end{picture}
\caption{\label{fig:dispcylRot} \small Dispersion of the $\langle
|a_{\ell m}|^2 \rangle^\Gamma$, normalized w.r.t. $C_\ell^\Gamma$, for low
multipoles ($2 \leq \ell \leq 5$), in a universe with cylindrical topology, as
a function of its orientation relative to the polar axis, and for different
values of the scale of compactification $L$.}
\end{figure*}

In order to show that this is a good measure of the alignment of multipoles,
and that it provides an efficient method to determine the directions of
possible alignments in real or simulated maps, let us compute the dispersion
of the squares $\langle |a_{\ell m}|^2 \rangle$, Eq. (\ref{Dispalm}), for a
cylinder which is oriented along a direction making an angle $\beta$ with the
$z$--axis. Each one of these squares can be computed with
\[
\langle |a_{\ell m}|^2 \rangle^{\Gamma}_R = \sum_{m_1} \left[ d_{mm_1}^\ell(\beta)
\right]^2 \langle |a_{\ell m_1}|^2\rangle^{\Gamma} \; ,
\]
which is nothing but (\ref{CorrMatCylRot}) restricted to the diagonal part.

In fig.\ref{fig:dispcylRot} it is shown this dispersion as a function of
$\beta$ for different multipole coefficients and for different values of $L$.
One can see that the dispersion has a maximum when the cylinder is oriented
along the $z$--axis. Thus in order to look for the alignments in a
hypothetical universe with the shape of a cylinder, one should just rotate the
celestial sphere around different directions until find those two opposite
ones along which the dispersion of the squares $|a_{\ell m}|^2$ is maximum.

However, in order to collect definitive evidence that the universe is indeed a
cylinder, one should map the dispersion of the squares $|a_{\ell m}|^2$ on the
sphere for each $\ell$--mode, i.e. one should determine the dispersion
(\ref{Dispalm}) as a function of the orientation of the celestial sphere. If
the universe had the topology of a cylinder, these dispersion maps should be
axially symmetric around a special direction, where the dispersion is maximum.
Moreover, this direction should be identified with the direction of
compactification of the cylinder.

If the universe has the topology of a flat homogeneous manifold note, from
(\ref{TopSignTorusDef}) and (\ref{TopSignTorusAux}), that the topological
signature is a superposition of rotated cylinders of different sizes. Thus a
CMB map for a universe with this kind of topology might present alignments
along the directions corresponding to these cylinders. In fact, rotating the
celestial sphere and computing the dispersion of the squares $|a_{\ell m}|^2$,
an easy computation shows that whenever we perform the rotation
$R(0,-\theta,-\varphi)$, with $\theta = \beta_{ij}$ and $\varphi =
\alpha_{ij}$, one has the cylinder labeled by $(i,j)$ oriented along the polar
axis, and thus \emph{dispersion maps might present local maxima along these
directions}.

Whether these local maxima are observable in a given dispersion map will
depend on (i) the scale of compactification of the corresponding cylinder
$L_i$, (ii) the background due to the simply connected part, and (iii) the
other cylinders' topological signatures. For large values of $L_i$ the
corresponding local maxima will not be observable, however one can expect
those maxima corresponding to the smaller cylinders to be detectable. The
existence and distribution of these maxima in each dispersion map, together
with their relative intensities is what we call a \emph{pattern of alignment}.

It might seem that the problem of constructing dispersion maps for manifolds
that are not flat homogeneous is more involved, since general cyclic manifolds
do not have axial symmetry as the cylinder has. Eq. (\ref{CorrMatCylRot})
depends on two angles only because the cylinder is axially symmetric, but in
the general case the expression for the correlation matrix in a rotated frame
depends on the three Euler angles. Thus it seems at first sight that in these
cases, a dispersion map should be a function on the 3--sphere. Fortunately,
the diagonal elements of the rotated correlation matrix depend only on the
last two Euler angles as
\begin{eqnarray*}
\langle |a_{\ell m}|^2 \rangle^{\Gamma}_R & = & \sum_{m_1,m_2} e^{i(m_2 -
m_1)\gamma} \times \\
& & \hspace{1.5cm} d_{mm_1}^\ell(\beta) d_{mm_2}^\ell(\beta) \times \\
& & \hspace{3.5cm} \langle a_{\ell m_1} a_{\ell m_2}^* \rangle^{\Gamma} \; ,
\end{eqnarray*}
thus the same conclusion holds in the general case. \emph{Dispersion maps on
the 2--sphere for low $\ell$--modes should display patterns of alignment
showing the symmetries of our Universe if it has a (not too large) nontrivial
topology}.

%%%%%%%%%%%%%%%%%%%%%%%%%%%%%%%%%%%%%%%%%%%%%%%%%%%%%%%%%%%%%%%%%%%%%%%%%%%%%%%
\section{Discussion}
\label{S:Discuss}
%%%%%%%%%%%%%%%%%%%%%%%%%%%%%%%%%%%%%%%%%%%%%%%%%%%%%%%%%%%%%%%%%%%%%%%%%%%%%%%

In order to study systematically the effects of a nontrivial spatial topology
in the temperature fluctuations of the CMB, we need to have the ability to
simulate efficiently temperature maps in multiply connected $\Lambda$CDM
cosmologies. Almost all the usual methods to perform these simulations use
explicitly the solutions of the Helmholtz equation in 3--manifolds with
nontrivial topology. The computation of the eigenfunctions and eigenvalues of
the Laplacian operator is simple only in Euclidean manifolds, while in
spherical and hyperbolic spaces it is a nontrivial problem. In fact, it is
only recently that an analytical computation has been achieved for all the
spherical manifolds. The hyperbolic cases still have to be done numerically.

In this paper we have developed a simulation procedure that avoids the
explicit use of the solutions of the Helmholtz equation. Instead, our results
are expressed in terms of the covering group $\Gamma$ of the corresponding
manifold. In this section we summarize the details of the method, its
efficiency, the simple applications performed here, and discuss future related
work.

%%%%%%%%%%%%%%%%%%%%%%%%%%%%%%%%%%%%%%%%%%%%%%%%%%%%%%%%%%%%%%%%%%%%%%%%%%%%%%%
\subsection{The formalism}
\label{Ss:Formal}
%%%%%%%%%%%%%%%%%%%%%%%%%%%%%%%%%%%%%%%%%%%%%%%%%%%%%%%%%%%%%%%%%%%%%%%%%%%%%%%

The cornerstone of our method is formula (\ref{two-point}), which is the
two--point correlation function of the scalar perturbations in a multiply
connected universe expressed in terms of the covering group of the manifold
\cite{BPS}. By means of simple formal manipulations we obtain an expression
for the correlation matrix of the spherical harmonic expansion coefficients of
the temperature maps, Eqs. (\ref{MCCorr})--(\ref{DefTopSign}), which contain
all the topological information expressed as a sum over the elements of the
covering group.

Former applications of (\ref{two-point}) required a regularization procedure
in order to account for divergences of the series, as well as some resummation
techniques for accelerating the convergence. We do not have these problems
here because the divergent series, which are actually distributions, appear
only inside integrals. Indeed, on the one hand, we show in Appendix
\ref{S:OldRes} that our formalism easily reproduces results previously
reported in the literature, as well as some simple generalizations, without
the need of any regularization procedure. On the other hand, elementary
decompositions of the two--point correlation function (\ref{two-point}), shown
in Sec. \ref{S:CycDec}, guarantee that our final expressions are highly
convergent, as discussed below.

Two decompositions of a generic covariance function which can be written as a
sum over the covering group are crucial for the efficiency of our formalism.
The first one, a trivial decomposition given by (\ref{TopSignGen}), defines
the topological signature of the covariance function. When written for the
correlation matrix of the harmonic expansion coefficients, it yields the
topological signature in the temperature anisotropy maps, as illustrated for
the cylinder by (\ref{CorrMatCyl}). This expression shows that the topological
signature is nothing but a ``perturbation'' of the correlation matrix
corresponding to the simply connected case. Since, as discussed in Sec.
\ref{Ss:Cylinder}, these ``perturbations'' are small, the efficiency of the
calculation follows.

The second decomposition given by (\ref{DecTSCyclic}) allows us to write the
topological signature of any manifold in terms of the topological signatures
of its maximal covering manifolds with cyclic covering groups. The example of
the tori illustrates the power of this approach, since we can write down
explicit formulae for a general torus whether its generating translations are
orthogonal and/or equal. Trying to do this with the explicit use of
eigenfunctions of the Laplacian (or with the method used in Appendix
\ref{S:OldRes}) turns out to be tedious if not difficult.

Moreover, by construction, this decomposition is invariant under the
symmetries of the manifold, thus it carries information on how shall these
symmetries manifest in individual CMB temperature anisotropy maps, as will be
discussed in Sec. \ref{Ss:TopSign}.

Another advantage of this second decomposition is the simplicity for writing
down the power spectrum for complicated manifolds. Expressions like
(\ref{ClTorus}) and (\ref{PowSpecChim}) are computationally very efficient
once we have saved the power spectrum for cyclic manifolds as a function of
its scale of compactification, since we have just to perform a weighted sum of
power spectra for cyclic manifolds at different scales considering the
multiplicity of the decomposition.

%%%%%%%%%%%%%%%%%%%%%%%%%%%%%%%%%%%%%%%%%%%%%%%%%%%%%%%%%%%%%%%%%%%%%%%%%%%%%%%
\subsection{Topological signatures}
\label{Ss:TopSign}
%%%%%%%%%%%%%%%%%%%%%%%%%%%%%%%%%%%%%%%%%%%%%%%%%%%%%%%%%%%%%%%%%%%%%%%%%%%%%%%

A further advantage of splitting the correlation matrix of the multipole
coefficients into its simply connected part and its topological signature is 
that we can identify very easily the geometric features of the signature.
Although we have made explicit calculations for flat homogeneous manifolds
only, qualitatively these results are general.

Universes with cylindrical topology of size $L \approx 2$ present clear
alignments of their low $\ell$--modes along the direction of compactification.
A dispersion map of the squares $\langle |a_{\ell m}|^2 \rangle$, for a given
low $\ell$, exhibits an axial symmetry around this direction, thus it reduces
to a function of the polar angle. These dispersion maps are shown in
fig.\ref{fig:dispcylRot}.

By decomposing the covering group $\Gamma$ in cyclic subgroups one can see
that, whatever the shape of our Universe, and if it is not too large,
dispersion maps (one for each individual low multipole) might show patterns of
alignment. In the general case such maps are functions on the two--dimensional
projective space or, by a lifting, on the 2--sphere. Although we have shown
the existence of patterns of alignment explicitly only for homogeneous flat
manifolds, it follows from the exposition of the general formalism that the
same conclusions hold for any manifold of constant curvature. Thus, we propose
the construction of these dispersion maps in the WMAP data, and so the search
for patterns of alignment, as a new method for detecting a possible nontrivial
topology of our Universe.

It is interesting to comment on some features relating Levin and
collaborators' proposal of pattern formation in CMB temperature maps and the
results we present in this paper. The patterns proposed by Levin et al.
\cite{LevinB, LSS98} are due to individual eigenmodes ($k$--modes), the
patterns we have identified here are due to multipole modes ($\ell$--modes).
In either case the modes compete to form their patterns in a CMB temperature
map, however the observable modes in a map on the sphere are the latter, since
spherical harmonics form a base on the space of functions on the sphere. On
the other side, the association between real space perturbations and angular
temperature fluctuations requires some averaging over the $k$--modes
\cite{Inoue03}, thus these patterns appear mixed in a map and their
observation might demand more elaborated techniques.

%%%%%%%%%%%%%%%%%%%%%%%%%%%%%%%%%%%%%%%%%%%%%%%%%%%%%%%%%%%%%%%%%%%%%%%%%%%%%%%
\subsection{Further remarks and future research}
\label{Ss:Future}
%%%%%%%%%%%%%%%%%%%%%%%%%%%%%%%%%%%%%%%%%%%%%%%%%%%%%%%%%%%%%%%%%%%%%%%%%%%%%%%

The formalism we have developed in this paper reveals new insights on the
problem of characterizing the marks that topology leaves in CMB maps, and
opens up new possibilities for developing further methods for unvealing the
shape of our Universe. It makes explicit that the multipole alignments
observed in COBE and WMAP full sky CMB temperature maps may be a manifestation
of its global shape, provides details of the nature and features of these
alignments, and gives at least one methodology to test this hypothesis. As a
consequence, further work is much needed.

One line of further research is the implementation of our formalism in the
spherical and hyperbolic cases. One way to do this requires first to identify
the radial part of the fundamental solution (\ref{HelmRadAng}) of the Helmholtz
equation in the universal covering, as well as the analog of the ``plane wave
expansion'' solution (\ref{SolHelmEuc}) in these geometries, and to write the
expansion of the corresponding ``plane waves'' in spherical harmonics as in
(\ref{PlaneW}). The difficult part seems to be expressing the ``plane wave
expansion'' in a suitable form to reproduce the formal steps used in the
Euclidean case.

Moreover we have to compute the topological signature of all other cyclic
manifolds, in order to extend the computations to any quotient space that could
be a candidate for the shape of space. We also need to include acoustic
oscillations, and Doppler and finite width effects in $\Lambda$CDM models so
that we could determine the relevant angular scales in Cosmic Topology, i.e.
the angular scales at which the topological signatures appear. This is a
crucial step in order to confront quantitatively the theory with real CMB maps
in an efficient and rigurous way. An ultimate goal may be to implement all
this methodology in a software package for public use.

The identification of the ``topological signature'' of the correlation matrix
$\langle a_{\ell m} \, a^*_{\ell' m'} \rangle^{\Gamma}$ also opens up a path
for solving a problem raised by Riazuelo et al. in \cite{RULW04}. The
correlation matrix for a multiply connected universe is non diagonal and,
tipically, $m$--dependent. In fact, this is the source of the statistical
anisotropy of the CMB in these universes. However, for very large manifolds
this correlation matrix becomes effectively diagonal, and equal to that
corresponding to the universal covering counterpart. A natural question raises
up: at what typical scales does the correlation matrix ``becomes diagonal''?
In terms of our formalism this problem can be stated as finding the scales at
where the topological signature becomes observationally negligible compared to
the simply connected part. A closed analysis of the topological signature
might give some answers to this and related questions. For example,
establishing bounds on the integral in (\ref{TopSignCyl}) might solve the
problem for the cylinder.

\vspace{3mm}
%%%%%%%%%%%%%%%%%%%%%%%%%%%%%%%%%%%%%%%%%%%%%%%%%%%%%%%%%%%%%%%%%%%%%%%%%%%%%%%
\section*{Acknowledgments}
%%%%%%%%%%%%%%%%%%%%%%%%%%%%%%%%%%%%%%%%%%%%%%%%%%%%%%%%%%%%%%%%%%%%%%%%%%%%%%%

We wish to thank Cristiane Camilo Hernandez for her unvaluable help with
fig.\ref{fig:mapas}, Wanderson Wanzeller for his continuous help in
computational issues, and Carlos Alexandre Wensche for showing us the papers
\cite{Aligne1} which triggered our interest in this topic. We would also like
to thank the Brazilian federal institutions CBPF and INPE for warm hospitality
in several ocassions, and to the participants of the Seminar of Cosmic
Topology held monthly at IFT and the Workshop New Physics from Space held
every year in Campos do Jord\~ao, S\~ao Paulo, where we had lots of
opportunities to discuss this work at the several stages of its development
during the last two years. W.S. Hip\'olito--Ricaldi acknowledges CAPES and
G.I. Gomero aknowledges FAPESP (contract 02/12328-6) for financial support.

\appendix

%%%%%%%%%%%%%%%%%%%%%%%%%%%%%%%%%%%%%%%%%%%%%%%%%%%%%%%%%%%%%%%%%%%%%%%%%%%%%%%
\section{Spherical harmonics}
\label{Ap:SphHarm}
%%%%%%%%%%%%%%%%%%%%%%%%%%%%%%%%%%%%%%%%%%%%%%%%%%%%%%%%%%%%%%%%%%%%%%%%%%%%%%%

In order to be self contained and to set the notation used in the paper, in
this appendix we present basic definitions, some useful formulae of spherical
harmonic functions and Wigner rotation matrices, as well as some invariance
properties of the correlation matrix $\langle a_{\ell m} \, a^*_{\ell' m'}
\rangle$ under coordinate transformations. For a complete treatment of
spherical harmonic functions the reader can consult \cite{VMK}.

%%%%%%%%%%%%%%%%%%%%%%%%%%%%%%%%%%%%%%%%%%%%%%%%%%%%%%%%%%%%%%%%%%%%%%%%%%%%%%%
\subsection{Basic definitions}
\label{Ss:BasicDef}
%%%%%%%%%%%%%%%%%%%%%%%%%%%%%%%%%%%%%%%%%%%%%%%%%%%%%%%%%%%%%%%%%%%%%%%%%%%%%%%

Let us denote by $\black{n} = (\theta,\varphi)$ a point in a 2--sphere
parametrized in the usual spherical coordinates, then the spherical harmonic
functions are defined as
\[
Y_{\ell m}(\black{n}) = \sqrt{\frac{2\ell + 1}{4\pi} \, \frac{(\ell - m)!}{(\ell + m)!}} \,
P_\ell^m(\cos \theta) \, e^{im \varphi} \; ,
\]
where
\[
P_\ell^m(x) = (-1)^m \, \left( 1-x^2 \right)^{m/2} \frac{d^m}{dx^m} \, P_\ell(x)
\]
are the associated Legendre functions with non--negative index $0 \leq m \leq
\ell$, and with
\[
P_\ell(x) = \frac{1}{2^\ell \, \ell !} \, \frac{d^\ell}{dx^\ell} \, \left( x^2
- 1 \right)^\ell
\]
being the Legendre polynomials. The associated Legendre functions with
negative index $m$ are defined by
\[
P_\ell^{-m}(x) = (-1)^m \, \frac{(\ell - m)!}{(\ell + m)!} \, P_\ell^m(x) \; .
\]
Moreover, it is often convenient to introduce the normalized associated
Legendre functions
\[
\mathcal{P}_\ell^m(x) = \sqrt{\frac{2\ell + 1}{2} \, \frac{(\ell - m)!}{(\ell + m)!}} \,
P_\ell^m(x) \; .
\]

It can easily be seen that the Legendre polynomial $P_\ell(x)$ is an $\ell$--th
degree polynomial of parity $\ell$, and thus the associated Legendre function
$P_\ell^m(x)$ is a function of parity $\ell - m$. It follows that the function
$\mathcal{P}_{\ell \ell' m}(x)$ defined in (\ref{PllmDef}) is an $(\ell +
\ell')$--th degree polynomial of parity $\ell + \ell'$, and thus the expression
for $F_{\ell \ell'}^m(x)$ in (\ref{CylFllmInit}), which is evaluated in
Appendix \ref{Ap:EvalFllm}, contains only even polynomials.

Spherical harmonics form a complete orthonormal set of functions on the
sphere, thus their most common application is in the expansion of functions,
like a CMB temperature anisotropy map, in multipoles as in
(\ref{TempSphHarm}), where the coefficients $a_{\ell m}$, called the multipole
coefficients, are given by
\[
a_{\ell m} = \int_{\mathbb{S}^2} d \Omega \, \frac{\delta T}{T}(\black{n}) \,
Y_{\ell m}^*(\black{n}) \; .
\]
Since the temperature map is a real function on the sphere, the multipole
coefficients obey the constraint $a_{\ell m}^* = (-1)^m \, a_{\ell , -m}$.

A very useful formula is given by the Addition Theorem for Spherical Harmonics
\begin{equation}
\label{AddFormula}
P_\ell (\black{n} \cdot \black{n}') = \frac{4\pi}{2 \ell + 1} \sum_{m = -
\ell}^\ell Y_{\ell m}(\black{n}) Y_{\ell m}^*(\black{n}') \; ,
\end{equation}
which for the particular case $\black{n} = \black{n}'$ yields the identity
\[
\sum_{m = - \ell}^\ell \left[ \mathcal{P}_\ell^m(x) \right]^2 = \frac{2 \ell +
1}{2} \; .
\]

%%%%%%%%%%%%%%%%%%%%%%%%%%%%%%%%%%%%%%%%%%%%%%%%%%%%%%%%%%%%%%%%%%%%%%%%%%%%%%%
\subsection{Wigner $D$--functions}
\label{Ss:Wigner}
%%%%%%%%%%%%%%%%%%%%%%%%%%%%%%%%%%%%%%%%%%%%%%%%%%%%%%%%%%%%%%%%%%%%%%%%%%%%%%%

In several ocassions it is convenient to rotate the sphere and compute the
multipole coefficients in this new coordinate system. This can be achieved by
means of the Wigner $D$--functions which can be defined operationally as
the functions $D_{mm_1}^\ell(R)$ such that, for any rotation $R \in SO(3)$, then
\[
Y_{\ell m}(R\black{n}) = \sum_{m_1} D_{mm_1}^\ell(R) \, Y_{\ell m_1}(\black{n}) \; .
\]
In this case, it can be shown that the multipole coefficients of the
temperature anisotropy map in the rotated reference frame are
\[
\widetilde{a}_{\ell m} = \sum_{m_1} D_{mm_1}^{*\ell}(R) \, a_{\ell m_1} \; .
\]
This expression can be used to compute the correlation matrix of the $a_{\ell
m}$'s in a rotated frame simply as
\begin{equation}
\label{a_lmRot}
\langle a_{\ell m} \, a^*_{\ell' m'} \rangle_R = \hspace{-0.2cm} \sum_{m_1,
m_1'} \hspace{-0.2cm} D_{mm_1}^{*\ell}(R) D_{m'm_1'}^{\ell'}(R) \langle
a_{\ell m_1} a^*_{\ell' m_1'} \rangle \; .
\end{equation}

The Wigner $D$--functions take a very simple form when we express the rotation
matrix $R$ in terms of its Euler angles as
\begin{equation}
\label{Euler}
R(\alpha, \beta, \gamma) = R_z(\alpha) \cdot R_y(\beta) \cdot R_z(\gamma) \; .
\end{equation}
Indeed, for this decomposition we have
\begin{equation}
\label{Wigner-Euler}
D_{mm'}^\ell (R(\alpha, \beta, \gamma)) = e^{i(m \alpha + m' \gamma)}
d_{mm'}^\ell(\beta) \; ,
\end{equation}
where $d_{mm'}^\ell(\beta) = D_{mm'}^\ell(R_y(\beta))$ is a real matrix with
the following symmetries
\begin{eqnarray*}
d_{mm'}^\ell(\beta) & = & (-1)^{m-m'} d_{m'm}^\ell(\beta) \; , \\
d_{mm'}^\ell(\beta) & = & d_{-m',-m}^\ell(\beta) \; , \\
d_{mm'}^\ell(\pi-\beta) & = & (-1)^{\ell - m'} d_{-m,m'}^\ell(\beta) \; , \\
d_{mm'}^\ell(-\beta) & = & (-1)^{m'-m} d_{m,m'}^\ell(\beta) \; .
\end{eqnarray*}
There exist several explicit and recursive formulae to compute these matrices
(see \cite{VMK}). A very efficient recursive procedure can be found in
\cite{BFB97}. The following formula will be enough to reproduce the results
presented in this paper.
\begin{widetext}
\[
d_{mm'}^\ell(\beta) = \sqrt{(\ell + m)!(\ell - m)!(\ell + m')!(\ell - m')!} \,
\sum_k{(-1)^k \, \frac{\left( \cos \frac{\beta}{2} \right)^{2\ell-2k+m-m'} \,
\left( \sin \frac{\beta}{2} \right)^{2k-m+m'}}{k! (\ell + m-k)! (\ell - m'-k)!
(m'-m+k)!}} \; ,
\]
\end{widetext}
where the sum in $k$ is evaluated whenever the arguments inside the factorials
are non--negative.

%%%%%%%%%%%%%%%%%%%%%%%%%%%%%%%%%%%%%%%%%%%%%%%%%%%%%%%%%%%%%%%%%%%%%%%%%%%%%%%
\subsection{Symmetry considerations}
\label{Ss:Symmetry}
%%%%%%%%%%%%%%%%%%%%%%%%%%%%%%%%%%%%%%%%%%%%%%%%%%%%%%%%%%%%%%%%%%%%%%%%%%%%%%%

Some consequences of the symmetries of the quotient manifold on the invariance
structure of the correlation matrix of the $a_{\ell m}$'s can be deduced
directly from the transformation rules of the spherical harmonic functions
under coordinate transformations. The results obtained in this way are formal,
generic, and are very useful in practical computations. We end this Appendix
by deducing the invariance properties the correlation matrix must have, given
some symmetries of the corresponding quotient manifold. These invariance
properties have been used in \cite{RULW04} to simplify the correlation matrix
for the 3--torus, however we want to remark here that they are general and do
not depend on the geometry of the universal covering space.

Let us begin with the invariance properties of $\langle a_{\ell m} \,
a^*_{\ell' m'} \rangle^{\Gamma}$ under rotations around the $z$--axis. Under a
rotation $R_z(\alpha): \varphi \to \varphi + \alpha$, the function $Y_{\ell
m}(\black{n})$ transforms as
\[
Y_{\ell m}(R_z(\alpha) \black{n}) = e^{im \alpha} Y_{\ell m}(\black{n}) \; .
\]
As a consequence the transformation rules for the multipole coefficients of a
CMB temperature map are of the form $\widetilde{a}_{\ell m} = e^{-im \alpha}
a_{\ell m}$, and so the correlation matrix transforms under this rotation as
\begin{equation}
\label{PhiTransfCorrMat}
\langle a_{\ell m} \, a^*_{\ell' m'} \rangle^{\Gamma}_{R_z(\alpha)} = e^{i(m'
- m) \alpha} \langle a_{\ell m} \, a^*_{\ell' m'} \rangle^{\Gamma} \; .
\end{equation}

We extract two consequences out of (\ref{PhiTransfCorrMat}). First, if the
quotient space is invariant under a rotation of $\alpha = 2 \pi/s$ around the
$z$--axis, then the correlation matrix must be zero unless $m = m'$ mod $s$.
Second, if the quotient space is invariant under ``any'' rotation around the
$z$--axis, the correlation matrix must be zero unless $m = m'$. In practice,
if we take our coordinate system such that the fundamental polyhedron of the
quotient manifold is oriented so that it is invariant under a $2\pi/s$
rotation around the polar axis, the correlation matrix will present a factor
$\delta_{m m'}^{\mbox{\tiny mod(s)}}$, and correspondingly, if the orientation
is such that the polyhedron is invariant under arbitrary rotations around the
$z$--axis, the correlation matrix will present a factor $\delta_{m m'}$.

Let us now take a look at invariance under the inversion transformation $P :
\black{n} \to - \black{n}$. Under this transformation the spherical harmonic
functions change as
\[
Y_{\ell m}(P \black{n}) = (-1)^{\ell} Y_{\ell m}(\black{n}) \; ,
\]
thus the multipole coefficients $a_{\ell m}$ change as $\widetilde{a}_{\ell m}
= (-1)^{\ell} a_{\ell m}$, and as a consequence the transformation rule for
the correlation matrix is
\begin{equation}
\label{InvTransfCorrMat}
\langle a_{\ell m} \, a^*_{\ell' m'} \rangle^{\Gamma}_P = (-1)^{\ell + \ell'}
\langle a_{\ell m} \, a^*_{\ell' m'} \rangle^{\Gamma} \; .
\end{equation}
Thus, if the fundamental polyhedron is oriented such that it appears invariant
under the parity transformation, the correlation matrix must be zero unless
$\ell = \ell'$ mod 2, i.e., the correlation matrix will present a factor
$\delta_{\ell \ell'}^{\mbox{\tiny mod(2)}}$.

To end this section, let us consider the reflection on the $y=0$ plane. This
operation changes only the azimuthal angle as $P_y : \varphi \to - \varphi$,
thus the transformation rule for the spherical harmonics are
\[
Y_{\ell m}(P_y \black{n}) = Y_{\ell m}^*(\black{n}) \; ,
\]
the multipole coefficients $a_{\ell m}$ change as $\widetilde{a}_{\ell m} =
a_{\ell m}^*$, and thus, the transformation rule for the correlation matrix is
\begin{equation}
\label{y0TransfCorrMat}
\langle a_{\ell m} \, a^*_{\ell' m'} \rangle^{\Gamma}_{P_y} = \langle a_{\ell
m} \, a^*_{\ell' m'} \rangle^{* \Gamma} \; .
\end{equation}
It immediately follows that if the fundamental polyhedron is oriented such
that it appears invariant under the reflection on the $y=0$ plane, the
correlation matrix must be real.

%%%%%%%%%%%%%%%%%%%%%%%%%%%%%%%%%%%%%%%%%%%%%%%%%%%%%%%%%%%%%%%%%%%%%%%%%%%%%%%
\section{Clausen functions}
\label{Ap:Clausen}
%%%%%%%%%%%%%%%%%%%%%%%%%%%%%%%%%%%%%%%%%%%%%%%%%%%%%%%%%%%%%%%%%%%%%%%%%%%%%%%

In this appendix we briefly present some computational aspects of the theory of
Clausen functions, as far as we need them for our purposes. Clausen functions
are periodic functions of period $2\pi$. There are two kinds of Clausen
functions, the $\varphi$--class and the $\psi$--class. Clausen
$\varphi$--functions can be expressed in terms of polynomials, while Clausen
$\psi$--functions involve higher transcendental functions, the so--called
Clausen integrals. Fortunately, we are interested exclusively in the Clausen
$\varphi$--functions, thus we will develop the details of the theory only for
them. The Clausen $\varphi$--functions are defined as
\begin{eqnarray}
\label{ClausenDefPhi}
\varphi_{2s-1}(x) & = & \sum_{n=1}^\infty \frac{\sin nx}{n^{2s-1}} \nonumber
\\ \\
\varphi_{2s}(x) & = & \sum_{n=1}^\infty \frac{\cos nx}{n^{2s}} \nonumber
\end{eqnarray}
for $s=1,2,\dots$, and can be calculated recursively with the formulae,
\begin{eqnarray}
\label{ClausenRecurse}
\varphi_{2s}(x) & = & \zeta(2s) - \int_0^x \varphi_{2s-1}(y) \, dy \nonumber
\\ \\
\varphi_{2s+1}(x) & = & \int_0^x \varphi_{2s}(y) \, dy \; , \nonumber
\end{eqnarray}
where
\[
\zeta(s) = \sum_{n=1}^\infty \frac{1}{n^s}
\]
is the Riemann Zeta function.

These recurrence relations are complemented by the initial condition
\begin{eqnarray}
\label{FirstClausen}
\varphi_1(x) & = & \sum_{n=1}^\infty \frac{\sin nx}{n} \nonumber \\
             & = & \frac{1}{2} \sum_{q \in \mathbb{Z}} \, [(2q+1)\pi -x]
\times \\
& & \hspace{1cm} \Theta(x - 2\pi q) \, \Theta(2\pi(q+1) - x) \; , \nonumber
\end{eqnarray}
where $\Theta(x)$ is the Heaviside step function. Formula (\ref{FirstClausen})
can be verified by computing the Fourier series of the second right hand side.

Since the Clausen functions are periodic of period $2 \pi$, we can write
\begin{eqnarray*}
\varphi_s(x) & = & \sum_{q \in \mathbb{Z}} \, f_s(x - 2 \pi q) \times \\
& & \hspace{1cm} \Theta(x - 2\pi q) \, \Theta(2\pi(q+1) - x) \; ,
\end{eqnarray*}
with $f_1(x) = \frac{1}{2} (\pi -x)$. The recurrence formulae
(\ref{ClausenRecurse}) yield the following expressions for the Clausen
functions in the period $[0,2\pi]$,
\begin{eqnarray}
\label{OddPerClaus}
f_{2s+1}(x) & = & \sum_{r=0}^{s-1} \frac{(-1)^r}{(2r+1)!} \, \zeta(2(s-r)) \,
x^{2r+1} + \\
& & \hspace{1.5cm} \frac{(-1)^s}{2} \left( \frac{\pi x^{2s}}{(2s)!} -
\frac{x^{2s+1}}{(2s+1)!} \right) \nonumber
\end{eqnarray}
for $s=0,1,2,\dots$, and
\begin{eqnarray}
\label{EvenPerClaus}
f_{2s}(x) & = & \sum_{r=0}^{s-1} \frac{(-1)^r}{(2r)!} \, \zeta(2(s-r)) \,
x^{2r} + \\
& & \hspace{1.5cm} \frac{(-1)^s}{2} \left( \frac{\pi x^{2s-1}}{(2s-1)!} -
\frac{x^{2s}}{(2s)!} \right) \nonumber
\end{eqnarray}
for $s=1,2,3,\dots$.

From the definitions (\ref{ClausenDefPhi}) we get $f_{2s+1}(\pi) = 0$, which
can be used to obtain a recurrence formula for the Riemann Zeta function of
even argument,
\begin{eqnarray}
\label{RiemannEven}
\zeta(2s) & = & \sum_{r=1}^{s-1} \frac{(-1)^{r+1}}{(2r+1)!} \, \zeta(2(s-r))
\, \pi^{2r} - \\
& & \hspace{3.5cm} \frac{(-1)^s \, s}{(2s+1)!} \, \pi^{2s} \; . \nonumber
\end{eqnarray}
Writing $\zeta(2s) = g_{2s}(0) \pi^{2s}$, and substituting this into
(\ref{RiemannEven}) we have
\[
g_{2s}(0) = \sum_{r=1}^{s-1} \frac{(-1)^{r+1}}{(2r+1)!} \, g_{2(s-r)}(0) -
\frac{(-1)^s \, s}{(2s+1)!} \; .
\]
The convenience for introducing this notation will be apparent in what follows.

We will now seek for generalizations of the formulae (\ref{OddPerClaus}) and
(\ref{EvenPerClaus}), i.e., we look for explicit expressions for the Clausen
functions in the $q$--th interval $[2\pi q, 2\pi (q+1)]$. Since the Clausen
functions satisfy the periodicity conditions $\varphi_{2s-1}(2\pi q) = 0$ and
$\varphi_{2s}(2\pi q) = \zeta(2s)$, the recurrence relations
(\ref{ClausenRecurse}) can be rewritten in the form
\begin{eqnarray}
\label{q-ClausenRecurse}
\varphi_{2s}(x) & = & \zeta(2s) - \int_{2\pi q}^x \varphi_{2s-1}(y) \, dy
\nonumber \\ \\
\varphi_{2s+1}(x) & = & \int_{2\pi q}^x \varphi_{2s}(y) \, dy \; , \nonumber
\end{eqnarray}

Defining the polynomials $f_s^q(x) = f_s(x-2\pi q)$,
we notice that $\varphi_s(x)$ coincides with $f_s^q(x)$ in the interval
$[2\pi q, 2\pi (q+1)]$. This fact, and the expressions
(\ref{q-ClausenRecurse}), allow us to write recurrence formulae analog to
(\ref{ClausenRecurse}) for the polynomials $f_s^q(x)$ as follows
\begin{eqnarray}
  f_{2s}^q(x) & = & g_{2s}(q) \pi^{2s} - \int_0^x f_{2s-1}^q(y) \, dy \; ,
\nonumber \\
\label{q-PerClausRecurse}
 & & \\
f_{2s+1}^q(x) & = & g_{2s+1}(q) \pi^{2s+1} + \int_0^x f_{2s}^q(y) \, dy \; ,
\nonumber
\end{eqnarray}
where
\begin{eqnarray}
  g_{2s}(q) & = & g_{2s}(0) + \frac{1}{\pi^{2s}} \int_0^{2\pi q} f_{2s-1}^q(y)
\, dy \; , \nonumber \\
\label{q-gRecurse}
 & & \\
g_{2s+1}(q) & = & - \frac{1}{\pi^{2s+1}} \int_0^{2\pi q} f_{2s}^q(y) \, dy
\; , \nonumber
\end{eqnarray}
with initial conditions, given by the first Clausen function, $f_1^q(x) =
g_1(q) \pi - \frac{x}{2}$ and $g_1(q) = q + \frac{1}{2}$.

The expressions (\ref{q-PerClausRecurse}) can be written in a unified way as
\[
f_s^q(x) = g_s(q) \pi^s - (-1)^s \int_0^x f_{s-1}^q(y) \, dy \; .
\]
Using this expression we readily obtain the explicit formula, which is the
generalization of (\ref{OddPerClaus}) and (\ref{EvenPerClaus}) we were looking
for,
\begin{eqnarray}
\label{q-PerClausRecurse2}
f_s^q(x) & = & \sum_{r=0}^{s-1} \frac{(-1)^{\mu(r,s)}}{r!} \, g_{s-r}(q)
\pi^{s-r} x^r - \\
& & \hspace{3.5cm} \frac{(-1)^{\mu(s,1)}}{2} \, \frac{x^s}{s!} \; , \nonumber
\end{eqnarray}
where
\[
\mu(r,s) = \left \lfloor \frac{r}{2} + \frac{1+ (-1)^s}{4} \right \rfloor \; ,
\]
and $\lfloor x \rfloor$ is the floor function of $x$, i.e., the largest
integer smaller than $x$.

The expressions (\ref{q-gRecurse}) can also be written in a unified way as
\[
g_s(q) = g_s(0) + \frac{(-1)^s}{\pi^s} \int_0^{2\pi q} f_{s-1}^q(y) \, dy \; ,
\]
where
\[
g_s(0) = \left\{ \begin{array}{l@{\qquad}l}
                    \frac{\zeta(s)}{\pi^s} & \mbox{if $s$ is even} \\
                    0             & \mbox{if $s>1$ is odd} \; .
                  \end{array} \right.
\]
From this we get the expression analogous to (\ref{q-PerClausRecurse2})
\begin{widetext}
\begin{equation}
\label{q-gRecurse2}
g_s(q) = g_s(0) + (-1)^s \left[ \sum_{r=1}^{s-1} (-1)^{\mu(r-1,s-1)} \,
\frac{2^r}{r!} \, g_{s-r}(q) \, q^r - (-1)^{\mu(s-1,1)} \, \frac{2^{s-1}}{s!}
\, q^s \right] \; .
\end{equation}
\end{widetext}

The polynomials $g_s(q)$ can also be written in the canonical form
\begin{equation}
\label{q-gExplicit}
g_s(q) = \sum_{k=0}^s A_k^s q^k \; ,
\end{equation}
where the coefficients are given by $A_0^s = g_s(0)$,
\[
A_n^s = (-1)^s \sum_{r=1}^n (-1)^{\mu(r-1,s-1)} \frac{2^r}{r!} \,
A_{n-r}^{s-r}
\]
for $0 < n < s$, and
\begin{eqnarray*}
A_s^s & = & (-1)^s \left[ \sum_{r=1}^{s-1} (-1)^{\mu(r-1,s-1)} \frac{2^r}{r!}
\, A_{s-r}^{s-r} \right. - \\
& & \hspace{3.5cm} \left. (-1)^{\mu(s-1,1)} \frac{2^{s-1}}{s!} \right] \; ,
\end{eqnarray*}
with initial conditions $A_0^1 = \frac{1}{2}$ and $A_1^1 = 1$. These
coefficients are obtained by just introducing (\ref{q-gExplicit}) into
(\ref{q-gRecurse2}) and collecting terms.

%%%%%%%%%%%%%%%%%%%%%%%%%%%%%%%%%%%%%%%%%%%%%%%%%%%%%%%%%%%%%%%%%%%%%%%%%%%%%%%
\section{The function $F_{\ell \ell'}^m(x)$}
\label{Ap:EvalFllm}
%%%%%%%%%%%%%%%%%%%%%%%%%%%%%%%%%%%%%%%%%%%%%%%%%%%%%%%%%%%%%%%%%%%%%%%%%%%%%%%

In this Appendix we evaluate the function $F_{\ell \ell'}^m(x)$ given by
(\ref{CylFllmInit}). We first observe (see Appendix \ref{Ap:SphHarm}) that the
function $\mathcal{P}_{\ell \ell' m}(x)$, given by (\ref{PllmDef}), is an even
polynomial of $(\ell + \ell')$--degree. Thus, we begin by considering the
integral
\[
I(\alpha) = \int_{-1}^1 P(y) \cos \alpha y \, dy \; ,
\]
where $P(y)$ is an even analytical function. Integrating succesively by parts
we get
\begin{eqnarray}
\label{AlphaInt}
I(\alpha) & = & 2 \left[ \frac{\sin \alpha}{\alpha} \sum_{s=0}^\infty
\frac{(-1)^s}{\alpha^{2s}} P^{(2s)}(1) + \right. \\
& & \hspace{1.5cm} \left. \frac{\cos \alpha}{\alpha^2}
\sum_{s=0}^\infty \frac{(-1)^s}{\alpha^{2s}} P^{(2s+1)}(1) \right] \; ,
\nonumber
\end{eqnarray}
where $P^{(k)}(x)$ is the $k$--th derivative of $P(x)$.

Making $\alpha = nx$ and $P(x) = \mathcal{P}_{\ell \ell' m}(x)$ in
(\ref{AlphaInt}), substituting (\ref{AlphaInt}) in (\ref{CylFllmInit}), and
performing the sum in $n$ we get
\begin{eqnarray}
\label{CylFllmClaus}
F_{\ell \ell'}^m(x) & = & 4 \sum_{s=0}^{\frac{\ell+\ell'}{2}} (-1)^s \left[
\frac{\mathcal{P}_{\ell \ell' m}^{(2s)}(1)}{x^{2s+1}} \, \varphi_{2s+1}(x)
\right. + \\
& & \hspace{2.5cm} \left. \frac{\mathcal{P}_{\ell \ell'
m}^{(2s+1)}(1)}{x^{2s+2}} \, \varphi_{2s+2}(x) \right] \; , \nonumber
\end{eqnarray}
where $\varphi_k(x)$ is the $k$--th Clausen $\varphi$--function defined in
Appendix \ref{Ap:Clausen}.

Since the Clausen functions are periodic functions of period $2\pi$, analytic
in each period, it follows that $F_{\ell \ell'}^m(x)$ is a piecewise continuous
function, analytic in each period as well. Thus we will now show how the
explicit expression for $F_{\ell \ell'}^m(x)$, in the $q$--th interval $[2\pi
q, 2\pi (q+1)]$, given in (\ref{PolyqCylFllmFin}), comes out.

Introducing the explicit form for the Clausen $\varphi$--functions
(\ref{q-PerClausRecurse2}), in the sum of (\ref{CylFllmClaus}) yields a huge
expresion, but a close inspection reveals that it is a polynomial in $\pi/x$.
The independent term is simply
\begin{eqnarray*}
- \frac{1}{2} \sum_{s=0}^{\frac{\ell+\ell'}{2}} \frac{(-1)^s}{(s+1)!} \,
\mathcal{P}_{\ell \ell' m}^{(s)}(1) & = & - \frac{1}{4} \int_{-1}^1
\mathcal{P}_{\ell \ell' m}(x) \, dx \\
 & = & - \frac{1}{4} \, \delta_{\ell \ell'} \; ,
\end{eqnarray*}
where the first equality can be deduced by writing the Taylor expansion of the
integrand of the right hand side, and integrating. On the other hand, summing
up all the coefficients of the $(r+1)$--th odd term, and proceeding as before,
we have this term equal to
\[
(-1)^r \, \mathcal{P}_{\ell \ell' m}^{(2r)}(0) \, g_{2r+1}(q) \left(
\frac{\pi}{x} \right)^{2r+1} \; ,
\]
while the $(r+1)$--th even term is equal to
\[
(-1)^r \, \mathcal{P}_{\ell \ell' m}^{(2r+1)}(0) \, g_{2r+2}(q) \left(
\frac{\pi}{x} \right)^{2r+2} \; ,
\]
which by the parity of $\mathcal{P}_{\ell \ell' m}(x)$ is zero. Summing up all the
terms we finally get (\ref{CylFllmFin}) and (\ref{PolyqCylFllmFin}).

%%%%%%%%%%%%%%%%%%%%%%%%%%%%%%%%%%%%%%%%%%%%%%%%%%%%%%%%%%%%%%%%%%%%%%%%%%%%%%%
\section{Known results for closed flat 3--manifolds}
\label{S:OldRes}
%%%%%%%%%%%%%%%%%%%%%%%%%%%%%%%%%%%%%%%%%%%%%%%%%%%%%%%%%%%%%%%%%%%%%%%%%%%%%%%

In this section we briefly show how we can obtain the formulae for the
correlation matrix of the $a_{\ell m}$'s and the angular power spectrum,
currently available in the literature, for some closed flat manifolds, as well
as a simple generalization, i.e., considering the observer out of the axis of
rotations of the screw motions of the covering group. We present explicit
derivations and formulae for the correlation matrix $\langle a_{\ell m}
a^*_{\ell' m'} \rangle$, as well as for the power spectrum, in order to allow
the interested reader to perform their own simulations confidently.

We first give a brief description of flat orientable closed 3--manifolds and
their covering groups. The versions of the diffeomorphic and isometric
classifications of flat 3--manifolds we present here were given by Wolf in
\cite{Wolf}, and previous descriptions in the context of cosmic topology were
given in \cite{Gomero} (see \cite{LevinA, RWULL04} for alternative
descriptions).

There are six diffeomorphic classes of compact orientable Euclidean
3--manifolds. The generators for the covering groups of the first five
classes, $\firstG-\fifthG$, are $\gamma_1 = (I, \black{a})$, $\gamma_2 = (I,
\black{b})$ and $\gamma_3 = (A_i, \black{c})$, where $A_1 = I$ is the identity
and
\begin{eqnarray*}
A_2 = \left( \begin{array}{ccc}
               -1 &  0 & 0 \\
               0 & -1 & 0 \\
               0 &  0 & 1
               \end{array} \right) & , &
A_4 = \left( \begin{array}{ccc}
              0 & -1 & 0 \\
              1 &  0 & 0 \\
              0 &  0 & 1
    \end{array} \right) \; , \\
A_3 = \left( \begin{array}{ccc}
              0 & -1 & 0 \\
              1 & -1 & 0 \\
              0 &  0 & 1
      \end{array} \right) & , &
A_5 = \left( \begin{array}{ccc}
              0 & -1 & 0 \\
              1 &  1 & 0 \\
              0 &  0 & 1
    \end{array} \right) \; ,
\end{eqnarray*}
for the classes $\firstG-\fifthG$ respectively. It is important to remark
that these matrices for the rotations are written in the basis formed by the
set $\{\black{a},\black{b},\black{c}\}$ of linearly independent vectors. Thus,
the torus $\firstG$ is generated by three independent translations, while for
the other manifolds the generators are two independent translations and a
screw motion along a linearly independent direction. The manifold $\sixthG$ is
the most involved since their generators are all screw motions. In the
following we present some general considerations concerning the classes
$\firstG-\fifthG$ only.

For space forms of the classes $\secondG-\fifthG$, the following facts are
easily derivable (see \cite{Gomero} for details):
\begin{enumerate}
\item The vector $\black{c}$ is orthogonal to both $\black{a}$ and
$\black{b}$.
\item \label{angle} The angle between $\black{a}$ and $\black{b}$ is a free
parameter for the class $\secondG$, while its value is f\/ixed to be $2\pi/3$,
$\pi/2$ and $\pi/3$ for the classes $\thirdG$, $\fourthG$ and $\fifthG$
respectively.
\item Denoting by $|\black{c}|$ the length of the vector $\black{c}$, and
similarly for any other vector, one has that $|\black{a}| = |\black{b}|$ for
the classes $\thirdG-\fifthG$, while both lengths are independent free
parameters in the class $\secondG$. Moreover, in all classes
$\secondG-\fifthG$, $|\black{c}|$ is an independent free parameter.
\item Denoting the canonical unitary basis vectors in Euclidean
space by $\{\base{x},\base{y},\base{z}\}$, one can always write $\black{a} =
|\black{a}| \, \base{x}$, $\black{b} = |\black{b}| \cos \varphi \, \base{x}
+ |\black{b}| \sin \varphi \, \base{y}$, and $\black{c} = |\black{c}| \,
\base{z}$, for the basis $\{\black{a},\black{b},\black{c}\}$, where $\varphi$
is the angle between $\black{a}$ and $\black{b}$, as established in the item
\ref{angle}.
\end{enumerate}

Thus in dealing with manifolds of classes $\secondG-\fifthG$, the axis of
rotation of the generator screw motion can be taken to be the $z$--axis, and
the orthogonal part of this generator, in the basis $\{\base{x}, \base{y},
\base{z}\}$, is
\begin{equation}
\label{Rot}
A = \left( \begin{array}{ccc}
         \cos \alpha & - \sin \alpha & 0 \\
         \sin \alpha &   \cos \alpha & 0 \\
          0      &        0      & 1
         \end{array} \right) \; ,
\end{equation}
with $\alpha=\pi$, $2\pi/3$, $\pi/2$ and $\pi/3$ respectively. Since the axis
of rotation passes through the origin, the translational part of the generator
$\gamma_3$ is $\black{c} = (0,0,L_z)$, where we have put $|\black{c}| = L_z$
as is usual in cosmic topology.

However, in cosmological applications we need to consider the arbitrariness of
the position of the observer inside space. Thus if the axis of rotation is at
a distance $\rho$ from the origin (the observer), and its intersection with
the horizontal plane makes an angle $\phi$ with the positive $x$--axis, the
translational part of the screw motion $\gamma_3 = (A,\black{c})$ is
\begin{eqnarray}
\label{transGen}
\black{c} & = & \rho [\cos \phi -\cos(\phi + \alpha)] \, \base{x} + \nonumber
\\
& & \hspace{0.3cm} \rho [\sin \phi - \sin(\phi + \alpha)] \, \base{y} + L_z
\, \base{z} \; .
\end{eqnarray}

In order to perform calculations of the topological signature of CMB
temperature maps, we need to write the covering group for the manifold under
study in a compact form. For a torus $\firstG$ the problem is trivial, since
the covering group is generated by three independent translations, and thus
any two isometries commute (see Sec.\ref{Ss:G1} below), while the covering
groups for the other closed flat manifolds are noncommutative since they
contain screw motions.

The generators of the covering groups for the classes $\secondG-\fifthG$
satisfy certain relations of the form
\[
\gamma_3 \gamma_1^{n_1} \gamma_2^{n_2} = \gamma_1^{m_1} \gamma_2^{m_2}
\gamma_3 \; ,
\]
where $n_1,n_2,m_1,m_2 \in \mathbb{Z}$, and they hold whether the axis of
rotation passes through the origin or not. It follows that a generic isometry
can always be put in the form
\begin{equation}
\label{GenIso}
\gamma = \gamma_1^{n_1} \gamma_2^{n_2} \gamma_3^{n_3} \; ,
\end{equation}
with $\gamma_1^{n_1} = (I,n_1 \black{a})$, $\gamma_2^{n_2} = (I,n_2
\black{b})$, and
\[
\gamma_3^{n_3} = (A^h, n_3 \black{c}_\parallel + \mathcal{O}_h
\black{c}_\perp) \; ,
\]
where $A$ is given by (\ref{Rot}), $\alpha = 2\pi/s$, $n_3 = sq+h$, with $q$
and $h$ integers such that $0 < h \leq s$, the parameter $s$ being $2,3,4$ and
6 corresponding to $\secondG$, $\thirdG$, $\fourthG$ and $\fifthG$
respectively,
\[
\oper_h = \sum_{j=0}^{h-1} A^j \; ,
\]
$\black{c}_\parallel = L_z \base{z}$, and $\black{c}_\perp = \rho [\cos \phi -
\cos(\phi + \alpha)] \, \base{x} + \rho [\sin \phi - \sin(\phi + \alpha)] \,
\base{y}$.

It is now straightforward to compute both the correlation matrix $\langle
a_{\ell m} a^*_{\ell' m'} \rangle$ and the angular power spectrum $C_\ell$.
They all have a simple structure. We first describe the general procedure for
obtaining these expressions and present the results in a unified form. We
finally specify each case separately. Note that, due to (\ref{transGen}), in
all of our calculations we are considering that the observer may be off an
axis of rotation of the screw motions of $\Gamma$.

Upon introducing (\ref{GenIso}) into (\ref{DefTopSign}), we transform the
series of exponentials in a series of Dirac's delta functions by using
(\ref{FirstId}). The integration of (\ref{MCCorr}) is then immediate in
Cartesian coordinates, the general result being
\begin{eqnarray*}
\langle a_{\ell m} a^*_{\ell' m'} \rangle & = & \frac{(4 \pi)^2}{V} \, i^{\ell
- \ell'} \sum_{\black{p} \in \widehat{\mathbb{Z}}^3} \frac{1}{\beta^3} \,
\Psi_{\ell \ell'}(2\pi \beta) \times \\
& & \hspace{1cm} Y_{\ell' m'}(\black{n}_{\vec{\beta}})
Y_{\ell m}^*(\black{n}_{\vec{\beta}}) \, f_{m'}^\Gamma(2\pi \vec{\beta}) \; ,
\end{eqnarray*}
where $V$ is the volume of the manifold, and $\widehat{\mathbb{Z}}^3 =
\mathbb{Z}^3 \setminus (0,0,0)$, since the term corresponding to $\black{p} =
0$ represents a constant perturbation, and thus is neglected.

The function
\[
f_{m}^\Gamma(\black{k}) = \frac{1}{s} \! \left[ 1 + \sum_{h=1}^{s-1}
\omega_s^{-hm} e^{-i \black{k} \cdot \oper_h \black{c}} \right] \; ,
\]
where $\omega_s$ is the first complex $s$th root of unity, is a complex
modulating term characteristic of the geometry and topology of the spatial
section of the universe model, and depends only on the screw motion
generators. The vector $\vec{\beta}(\black{p})$ comes from the discretization
of the wavevector $\black{k}$ due to the Dirac's deltas (each $2\pi \beta$ is
an eigenvalue of the Laplacian operator), and $\black{n}_{\vec{\beta}}$ is the
unit vector in the direction of $\vec{\beta}$.

Using the property $\langle a_{\ell m} a^*_{\ell' m'} \rangle = \langle
a_{\ell' m'} a^*_{\ell m} \rangle^*$ one can easily show, by resumming the
series, that the variances of the multipole moments can be put in the general
form
\begin{eqnarray*}
\langle |a_{\ell m}|^2 \rangle & = & \frac{(4\pi)^2}{V} \sum_{\black{p} \in
\widehat{\mathbb{Z}}^3} \frac{1}{\beta^3} \, \Psi_{\ell \ell}(2\pi \beta)
\times \\
& & \hspace{1.7cm} \left| Y_{\ell m}(\black{n}_{\vec{\beta}}) \right|^2 \,
\Re\left( f_{m}^\Gamma(\vec{2\pi \beta}) \right) \; ,
\end{eqnarray*}
where $\Re$ stands for the real part of a complex number. The angular power
spectrum is then
\[
C_\ell = \frac{4\pi}{V} \sum_{\black{p} \in \widehat{\mathbb{Z}}^3}
\frac{1}{\beta^3} \, \Psi_{\ell \ell}(2\pi \beta) \, \Xi_\ell(2\pi
\vec{\beta}) \; ,
\]
where
\[
\Xi_\ell(\black{k}) = \frac{4\pi}{2\ell + 1} \sum_{m=-l}^\ell \left|
Y_{\ell m}(\black{n}_{\black{k}}) \right|^2 \Re\left(
f_{m}^\Gamma(\black{k}) \right)
\]
can be evaluated using the Addition Theorem for Spherical Harmonics yielding
\[
\Xi_\ell(\black{k}) = \frac{1}{s} \left[ 1 + \sum_{h=1}^{s-1} P_\ell(\cos
\theta_{\black{k},h}) \cos(\black{k} \cdot \oper_h \black{c}) \right] \; ,
\]
where
\[
\cos \theta_{\black{k},h} = \cos^2 \theta_{\black{k}} + \sin^2
\theta_{\black{k}} \, \cos \frac{2\pi h}{s} \; .
\]

%%%%%%%%%%%%%%%%%%%%%%%%%%%%%%%%%%%%%%%%%%%%%%%%%%%%%%%%%%%%%%%%%%%%%%%%%%%%%%%
\subsection{Rectangular torus $\firstG$}
\label{Ss:G1}
%%%%%%%%%%%%%%%%%%%%%%%%%%%%%%%%%%%%%%%%%%%%%%%%%%%%%%%%%%%%%%%%%%%%%%%%%%%%%%%

The generators for the rectangular torus are the translations $\black{a} = L_x
\base{x}$, $\black{b} = L_y \base{y}$ and $\black{c} = L_z \base{z}$, thus a
generic isometry of its covering group can be written as $\gamma =
(I,\black{r})$, with $\black{r} = n_xL_x \base{x} + n_yL_y \base{y} + n_zL_z
\base{z}$, and $n_x,n_y,n_z \in \mathbb{Z}$, i.e., the covering group of
$\firstG$ is parametrized by $\mathbb{Z}^3$.

It follows immediately that, for a rectangular torus, the expression
(\ref{DefTopSign}) takes the form
\begin{eqnarray*}
\Upsilon^{\Gamma}_{\ell m} (\black{k}) & = & \sum_{\black{n} \in
\mathbb{Z}^3} \! e^{-i(n_xk_xL_x + n_yk_yL_y + n_zk_zL_z)}
Y_{\ell m}(\black{n}_\black{k}) \\
& = & (2 \pi)^3 \sum_{\black{p} \in \mathbb{Z}^3} \,
\delta(k_xL_x \! - \! 2 \pi p_x) \times \\
& &  \delta(k_yL_y \! - \! 2 \pi p_y) \,
\delta(k_zL_z \! - \! 2 \pi p_z) Y_{\ell m}(\black{n}_\black{k}) \; .
\end{eqnarray*}
Following the procedure described above one gets $f_{m}^\Gamma(\black{k}) =
1$, and $\beta_x = \frac{p_x}{L_x}$, $\beta_y = \frac{p_y}{L_y}$, and $\beta_z
= \frac{p_z}{L_z}$. In particular, we have the well known result
\[
C_\ell = \frac{4\pi}{V} \sum_{\black{p} \in \widehat{\mathbb{Z}}^3}
\frac{1}{\beta^3} \, \Psi_{\ell \ell}(2\pi \beta) \; .
\]

%%%%%%%%%%%%%%%%%%%%%%%%%%%%%%%%%%%%%%%%%%%%%%%%%%%%%%%%%%%%%%%%%%%%%%%%%%%%%%%
\subsection{Rectangular $\secondG$}
\label{Ss:G2}
%%%%%%%%%%%%%%%%%%%%%%%%%%%%%%%%%%%%%%%%%%%%%%%%%%%%%%%%%%%%%%%%%%%%%%%%%%%%%%%

The generators for the rectangular $\secondG$ are $\gamma_1 = (I, \black{a})$,
$\gamma_2 = (I, \black{b})$ and $\gamma_3 = (A, \black{c})$, with $\black{a}
= L_x \base{x}$, $\black{b} = L_y \base{y}$, $\black{c} = 2 \rho \cos \phi \,
\base{x} + 2 \rho \sin \phi \, \base{y} + L_z \base{z}$, and $A$ given in
(\ref{Rot}) with $\alpha = \pi$. They satisfy the relations $\gamma_1 \gamma_3
\gamma_1 = \gamma_3$ and $\gamma_2 \gamma_3 \gamma_2 = \gamma_3$, which allow
to write any isometry of the covering group by (\ref{GenIso}) with
\begin{equation}
\label{G2gamma3}
\gamma_3^{n_3} = \left\{ \begin{array}{lcc}
      (I, n_3 \black{c}_\parallel) & & \mbox{if $n_3$ is even} \\
(A, n_3 \black{c}_\parallel + \black{c}_\perp) & & \mbox{if $n_3$ is odd}
                         \end{array} \right. \; ,
\end{equation}
where $\black{c}_\perp = 2 \rho \cos \phi \, \base{x} + 2 \rho \sin \phi \,
\base{y}$.

It follows from (\ref{GenIso}) and (\ref{G2gamma3}) that the expression
(\ref{DefTopSign}) takes the form
\begin{widetext}
\begin{eqnarray*}
\Upsilon^{\Gamma}_{\ell m} (\black{k}) & = & \sum_{\black{n} \in
\mathbb{Z}^3} e^{-i(n_xk_xL_x + n_yk_yL_y + 2n_zk_zL_z)} \left[ 1 + (-1)^m
e^{-i \black{k} \cdot \black{c}} \right] \, Y_{\ell m}(\black{n}_\black{k}) \\
& = & (2 \pi)^3 \sum_{\black{p} \in \mathbb{Z}^3} \, \delta(k_xL_x - 2 \pi
p_x) \, \delta(k_yL_y - 2 \pi p_y) \, \delta(k_zL_z - \pi p_z) \,
Y_{\ell m}(\black{n}_\black{k}) \, f_{m}^\Gamma(\black{k}) \; ,
\end{eqnarray*}
\end{widetext}
where we have put $n_1 = n_x$, $n_2 = n_y$, and $n_3 = 2n_z$ or $2n_z + 1$,
depending on whether $n_3$ is even or odd. The components of $\vec{\beta}$
are $\beta_x = \frac{p_x}{L_x}$, $\beta_y = \frac{p_y}{L_y}$, and $\beta_z =
\frac{p_z}{2L_z}$.

%%%%%%%%%%%%%%%%%%%%%%%%%%%%%%%%%%%%%%%%%%%%%%%%%%%%%%%%%%%%%%%%%%%%%%%%%%%%%%%
\subsection{$\thirdG$}
\label{Ss:G3}
%%%%%%%%%%%%%%%%%%%%%%%%%%%%%%%%%%%%%%%%%%%%%%%%%%%%%%%%%%%%%%%%%%%%%%%%%%%%%%%

The generators for a manifold of class $\thirdG$ are $\gamma_1 = (I,
\black{a})$, $\gamma_2 = (I, \black{b})$, and $\gamma_3 = (A, \black{c})$,
with $\black{a} = L \, \base{x}$, $\black{b} = - \frac{L}{2} \, (\base{x} -
\sqrt{3} \, \base{y})$, $\black{c} = \frac{\rho}{2} \, (3 \cos \phi + \sqrt{3}
\sin \phi) \, \base{x} + \frac{\rho}{2} \, (3 \sin \phi - \sqrt{3} \cos \phi)
\, \base{y} + L_z \base{z}$, and $A$ given in (\ref{Rot}) with $\alpha =
2\pi/3$. They satisfy the relations $\gamma_2^{-1} \gamma_3 \gamma_1 =
\gamma_3$ and $\gamma_1 \gamma_2 \gamma_3 \gamma_2 = \gamma_3$, which allow us
to write any isometry of the covering group by (\ref{GenIso}) with
\begin{equation}
\label{G3gamma3}
\gamma_3^{n_3} = \left\{ \hspace{-0.1cm} \begin{array}{lcc}
(I, n_3 \black{c}_\parallel) & & \hspace{-0.2cm} \mbox{if } n_3 = 0 \mbox{ mod
3} \\
(A, n_3 \black{c}_\parallel + \black{c}_\perp) & & \hspace{-0.2cm} \mbox{if }
n_3 = 1 \mbox{ mod 3} \\
(A^2, n_3 \black{c}_\parallel + \oper_2 \black{c}_\perp) & & \hspace{-0.2cm}
\mbox{if } n_3 = 2 \mbox{ mod 3}
                         \end{array} \right. \; ,
\end{equation}
where $\black{c}_\perp = \frac{\rho}{2} \, (3 \cos \phi + \sqrt{3} \sin \phi)
\, \base{x} + \frac{\rho}{2} \, (3 \sin \phi - \sqrt{3} \cos \phi) \,
\base{y}$.

It follows from (\ref{GenIso}) and (\ref{G3gamma3}) that the expression
(\ref{DefTopSign}) takes the form
\begin{widetext}
\begin{eqnarray*}
\Upsilon^{\Gamma}_{\ell m} (\black{k}) & = & \sum_{\black{n} \in
\mathbb{Z}^3} e^{-i\left[n_xk_xL + n_y \left(\frac{\sqrt{3}}{2}k_y -
\frac{1}{2}k_x\right)L + 3n_zk_zL_z\right]} \left[ 1 + \omega_3^{-m} e^{-i
\black{k} \cdot \black{c}} + \omega_3^{-2m} e^{-i \black{k} \cdot \oper_2
\black{c}} \right] Y_{\ell m}(\black{n}_\black{k}) \\
& = & (2 \pi)^3 \sum_{\black{p} \in \mathbb{Z}^3} \, \delta(k_xL - 2 \pi p_x)
\, \delta \left( \left[ {\textstyle \frac{\sqrt{3}}{2}} k_y - {\textstyle
\frac{1}{2}} k_x \right]L - 2 \pi p_y \right) \, \delta \left( k_zL_z -
{\textstyle \frac{2\pi}{3}} \, p_z \right) Y_{\ell m}(\black{n}_\black{k}) \,
f_{m}^\Gamma(\black{k}) \; ,
\end{eqnarray*}
\end{widetext}
where we have put $n_1 = n_x$, $n_2 = n_y$, and $n_3 = 3n_z$, $3n_z + 1$ or
$3n_z + 2$ according to (\ref{G3gamma3}). We also get $\beta_x =
\frac{p_x}{L}$, $\beta_y = \frac{\sqrt{3}}{3L} \, (2p_y + p_x)$, and $\beta_z
= \frac{p_z}{3L_z}$.

%%%%%%%%%%%%%%%%%%%%%%%%%%%%%%%%%%%%%%%%%%%%%%%%%%%%%%%%%%%%%%%%%%%%%%%%%%%%%%%
\subsection{$\fourthG$}
\label{Ss:G4}
%%%%%%%%%%%%%%%%%%%%%%%%%%%%%%%%%%%%%%%%%%%%%%%%%%%%%%%%%%%%%%%%%%%%%%%%%%%%%%%

The generators for a manifold of class $\fourthG$ are $\gamma_1 = (I,
\black{a})$, $\gamma_2 = (I, \black{b})$, and $\gamma_3 = (A, \black{c})$,
with $\black{a} = L \, \base{x}$, $\black{b} = L \, \base{y}$, $\black{c} =
\rho \, (\cos \phi + \sin \phi) \, \base{x} + \rho \, (\sin \phi - \cos \phi)
\, \base{y} + L_z \base{z}$, and $A$ given in (\ref{Rot}) with $\alpha =
\pi/2$. They satisfy the relations $\gamma_2^{-1} \gamma_3 \gamma_1 =
\gamma_3$ and $\gamma_1 \gamma_3 \gamma_2 = \gamma_3$, which allow us to write
any isometry of the covering group by (\ref{GenIso}) with
\begin{equation}
\label{G4gamma3}
\gamma_3^{n_3} = \left\{ \begin{array}{lcc}
      (I, n_3 \black{c}_\parallel) & & \mbox{if } n_3 = 0 \mbox{ mod 4} \\
(A, n_3 \black{c}_\parallel + \black{c}_\perp) & & \mbox{if } n_3 = 1
\mbox{ mod 4} \\
(A^2, n_3 \black{c}_\parallel + \oper_2 \black{c}_\perp) & & \mbox{if }
n_3 = 2 \mbox{ mod 4} \\
(A^3, n_3 \black{c}_\parallel + \oper_3 \black{c}_\perp) & & \mbox{if }
n_3 = 3 \mbox{ mod 4}
                         \end{array} \right. \; ,
\end{equation}
where $\black{c}_\perp = \rho \, (\cos \phi + \sin \phi) \, \base{x} + \rho \,
(\sin \phi - \cos \phi) \, \base{y}$.

Similarly, using (\ref{GenIso}) and (\ref{G4gamma3}), the expression
(\ref{DefTopSign}) takes the form
\begin{widetext}
\begin{eqnarray*}
\Upsilon^{\Gamma}_{\ell m} (\black{k}) & = & \hspace{-0.1cm} \sum_{\black{n}
\in \mathbb{Z}^3} \hspace{-0.1cm} e^{-i (n_xk_xL + n_yk_yL + 4n_zk_zL_z)}
\left[ 1 + \sum_{h=1}^3 \omega_4^{-hm} e^{-i \black{k} \cdot \oper_h
\black{c}} \right] Y_{\ell m}(\black{n}_\black{k}\!) \\
& = & (2 \pi)^3 \sum_{\black{p} \in \mathbb{Z}^3} \, \delta(k_xL - 2 \pi p_x)
\, \delta (k_yL - 2 \pi p_y) \, \delta \left( k_zL_z - {\textstyle
\frac{\pi}{2}} \, p_z \right) \, Y_{\ell m}(\black{n}_\black{k}) \,
f_{m}^\Gamma(\black{k}) \; ,
\end{eqnarray*}
\end{widetext}
where we have put $n_1 = n_x$, $n_2 = n_y$, and $n_3 = 4n_z$, $4n_z + 1$,
$4n_z + 2$ or $4n_z + 3$ according to (\ref{G4gamma3}), and $\omega_4$ is the
first complex 4th-rooth of unity. We also get $\beta_x = \frac{p_x}{L}$,
$\beta_y = \frac{p_y}{L}$, and $\beta_z = \frac{p_z}{4L_z}$.

%%%%%%%%%%%%%%%%%%%%%%%%%%%%%%%%%%%%%%%%%%%%%%%%%%%%%%%%%%%%%%%%%%%%%%%%%%%%%%%
\subsection{$\fifthG$}
\label{Ss:G5}
%%%%%%%%%%%%%%%%%%%%%%%%%%%%%%%%%%%%%%%%%%%%%%%%%%%%%%%%%%%%%%%%%%%%%%%%%%%%%%%

The generators for the rectangular $\fifthG$ are $\gamma_1 = (I, \black{a})$,
$\gamma_2 = (I, \black{b})$, and $\gamma_3 = (A, \black{c})$, with $\black{a}
= L \, \base{x}$, $\black{b} = \frac{L}{2} \, (\base{x} + \sqrt{3} \,
\base{y})$, $\black{c} = \frac{\rho}{2} \, (\cos \phi + \sqrt{3} \sin \phi) \,
\base{x} + \frac{\rho}{2} \, (\sin \phi - \sqrt{3} \cos \phi) \, \base{y} +
L_z \base{z}$, and $A$ given in (\ref{Rot}) with $\alpha = \pi/3$. They
satisfy the relations $\gamma_2^{-1} \gamma_3 \gamma_1 = \gamma_3$ and
$\gamma_1 \gamma_2^{-1} \gamma_3 \gamma_2 = \gamma_3$, which allow us to write
any isometry of the covering group by (\ref{GenIso}) with
\begin{equation}
\label{G5gamma3}
\gamma_3^{n_3} = \left\{ \begin{array}{lcc}
      (I, n_3 \black{c}_\parallel) & & \mbox{if } n_3 = 0 \mbox{ mod 6} \\
(A, n_3 \black{c}_\parallel + \black{c}_\perp) & & \mbox{if } n_3 = 1
\mbox{ mod 6} \\
(A^2, n_3 \black{c}_\parallel + \oper_2 \black{c}_\perp) & & \mbox{if }
n_3 = 2 \mbox{ mod 6} \\
(A^3, n_3 \black{c}_\parallel + \oper_3 \black{c}_\perp) & & \mbox{if }
n_3 = 3 \mbox{ mod 6} \\
(A^4, n_3 \black{c}_\parallel + \oper_4 \black{c}_\perp) & & \mbox{if }
n_3 = 4 \mbox{ mod 6} \\
(A^5, n_3 \black{c}_\parallel + \oper_5 \black{c}_\perp) & & \mbox{if }
n_3 = 5 \mbox{ mod 6}
                         \end{array} \right. \; ,
\end{equation}
where $\black{c}_\perp = \frac{\rho}{2} \, (\cos \phi + \sqrt{3} \sin \phi) \,
\base{x} + \frac{\rho}{2} \, (\sin \phi - \sqrt{3} \cos \phi) \, \base{y}$.

Using (\ref{GenIso}) and (\ref{G5gamma3}), the expression (\ref{DefTopSign})
takes the form
\begin{widetext}
\begin{eqnarray*}
\Upsilon^{\Gamma}_{\ell m} (\black{k}) & = & \sum_{\black{n} \in \mathbb{Z}^3}
e^{-i\left[n_xk_xL + n_y \left(\frac{\sqrt{3}}{2}k_y + \frac{1}{2}k_x\right)L
+ 6n_zk_zL_z\right]} \left[ 1 + \sum_{h=1}^5 \omega_6^{-hm} e^{-i \black{k}
\cdot \oper_h \black{c}} \right] Y_{\ell m}(\black{n}_\black{k}) \\
& = & (2 \pi)^3 \sum_{\black{p} \in \mathbb{Z}^3} \, \delta(k_xL - 2 \pi p_x)
\, \delta \left( \left[ {\textstyle \frac{\sqrt{3}}{2}} k_y + {\textstyle
\frac{1}{2}} k_x \right]L - 2 \pi p_y \right) \, \delta \left( k_zL_z -
{\textstyle \frac{\pi}{3}} \, p_z \right) Y_{\ell m}(\black{n}_\black{k}) \,
f_{m}^\Gamma(\black{k}) \; ,
\end{eqnarray*}
\end{widetext}
where we have put $n_1 = n_x$, $n_2 = n_y$, and $n_3 = 6n_z$, $6n_z + 1$,
$6n_z + 2$, $\dots$, $6n_z + 5$ according to (\ref{G5gamma3}), and $\omega_6$
is the first complex 6th-rooth of unity. We also get $\beta_x =
\frac{p_x}{L}$, $\beta_y = \frac{\sqrt{3}}{3L} \, (2p_y - p_x)$, and $\beta_z
= \frac{p_z}{6L_z}$.

%%%%%%%%%%%%%%%%%%%%%%%%%%%%%%%%%%%%%%%%%%%%%%%%%%%%%%%%%%%%%%%%%%%%%%%%%%%%%%%

%%%%%%%%%%%%%%%%%%%%%%%%%%%%%%%%%%%%%%%%%%%%%%%%%%%%%%%%%%%%%%%%%%%%%%%%%%%%%%%
%
%
\end{document}